\documentclass[11pt]{article}
\usepackage{amsmath}
\usepackage{amsfonts}
\usepackage{amssymb}
\usepackage{amscd}
\usepackage{epsfig}
\usepackage{times}
\newcommand{\ol}{\overline}
\newcommand{\bs}{\boldsymbol}

\renewcommand{\l}{\newline\null}
\begin{document}

\begin{titlepage}
%
%\today\quad(\printtime)\hfill PAR-LPTHE 03/xx
January 2003\hfill PAR-LPTHE 03/05
%\begin{flushright} hep-ph/0301274 \end{flushright}
%\vskip 3.5cm
\vskip 4cm
{\baselineskip 17pt
%\begin{center}
\centerline{\bf ABOUT NEUTRAL KAONS AND SIMILAR SYSTEMS;}
\centerline{\bf FROM QUANTUM FIELD THEORY TO EFFECTIVE MASS MATRICES}
%\end{center}
}
\vskip .5cm
\centerline{B. Machet
     \footnote[1]{Member of `Centre National de la Recherche Scientifique'}
     \footnote[2]{E-mail: machet@lpthe.jussieu.fr}
     }
\vskip 5mm
\centerline{{\em Laboratoire de Physique Th\'eorique et Hautes \'Energies}
     \footnote[3]{LPTHE tour 16, 1\raise 3pt \hbox{\tiny er} \'etage,
          Universit\'e P. et M. Curie, BP 126, 4 place Jussieu,
          F-75252 Paris Cedex 05 (France)}
}
\centerline{\em Universit\'es Pierre et Marie Curie (Paris 6) et Denis
Diderot (Paris 7)}
\centerline{\em Unit\'e associ\'ee au CNRS UMR 7589}
\vskip 1cm
{\bf Abstract:} Systems of neutral interacting mesons are investigated,
concerning in particular the validity of their description by an
effective hamiltonian.
First, I study its connection to quantum field theory and show
that the spectrum of such systems  cannot be reduced in
general to the one of a single constant effective mass matrix.
Choosing nevertheless to work in this customary formalism,
one then faces several ways to diagonalize a complex matrix, which
lead to different eigenvalues and eigenvectors.
Last, and it is the main subject of this work, because $K^0$ and its
charge conjugate $\ol{K^0}$ are also connected, in quantum field theory, by
hermitian conjugation, any constant effective mass matrix is defined, in
this basis, up to arbitrary diagonal antisymmetric terms; I use this
freedom to  deform the mass matrix in various ways and study the
consequences on its spectrum.
Emphasis is put on the role of discrete symmetries throughout the paper.
That the  degeneracy of the eigenvalues of the full renormalized
mass matrix can be a sufficient condition for the outcome of $CP$ violation
is outlined in an appendix.
In the whole work, the dual formalism of $|\ in>$ and $<out\ |$ states
and bi-orthogonal basis, suitable for non-normal matrices, is used.
\smallskip


\vfill
%\null\hfil\epsffile{/users/lpthe/machet/Papiers/Logo/LogoCNRS.ps}
\null\hfil\epsffile{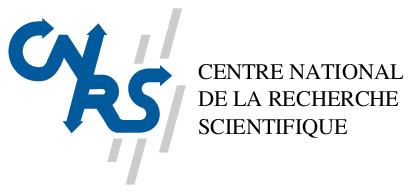}
\end{titlepage}
%
%%%%%%%%%%%%%%%%%%%%%%%%%%%%%%%%%%%%%%%%%%%%%%%%%%%%%%%%%%%%%%%%%%%%%%%%%%%%%%
%%%%%%%%%%%%%%%%%%%%%%%%%%%%%%%%%%%%%%%%%%%%%%%%%%%%%%%%%%%%%%%%%%%%%%%%%%%%%%

%%%%%%%%%%%%%%%%%%%%%%%%%%%%%%%%%%%%%%%%%%%%%%%%%%%%%%%%%%%%%%%%%%%%%%%%%%%%%%
\section{INTRODUCTION}
\label{section:introduction}
%%%%%%%%%%%%%%%%%%%%%%%%%%%%%%%%%%%%%%%%%%%%%%%%%%%%%%%%%%%%%%%%%%%%%%%%%%%%%%
%
We face two pictures of neutral kaons: flavour (strangeness)
eigenstates $(K^0,\ol{K^0})$ are concerned with electroweak -- for example
semi-leptonic -- decays,
 and $(K_L,K_S)$ with decays where strong interactions are also involved
 (for example in the disintegrations into two or three pions).

On one side, the particle-antiparticle  flavour eigenstates
are constrained to have the same mass by $CPT$
symmetry and, on the other side, the $K_L-K_S$ mass difference stands
among the quantities measured with the highest precision today.
So, whether there could exist a twofold (at least)  set of mass
eigenstates for such systems is worth investigating;
which one is detected depends on
which type of decay is analyzed and which  quantum numbers are observed.

Along this line, the question naturally arises of the uniqueness
of the effective mass matrix $M$ attached to such systems.  This
paper  points at  ambiguities in the definition of $M$ and
eventually uses them to ``deform'' it.

Since the system under scrutiny is unstable, its mass matrix must be
taken non-hermitian, and it is  necessary to use the formalism
of $|\ in>$ and $<out\ |$ states  \cite{BrancoLavouraSilva}\cite{Silva};
the latter only coincide when the mass matrix is normal ($CP$ conserved).
However, in all cases, the same $(K^0,\ol{K^0})$ basis can be chosen
to span both Hilbert
spaces ${\cal V}_{in}$ and ${\cal V}_{out}$ which accordingly can
be identified.

The second section is devoted to a survey of the uncertainties and
ambiguities which occur when going from  quantum field theory (QFT) to an
effective hamiltonian
formalism for the determination of the masses and mass eigenstates
of a system of interacting mesons. It is shown that, in general, the
the problem of finding its mass spectrum and eigenstates cannot  be reduced to
the diagonalization of a single constant mass matrix. Emphasis is put on
the role of discrete symmetries.

Despite this restriction, the rest of the paper is set in the formalism of
a single constant effective mass matrix.

The third section is devoted to the evaluation of
$|\ K^0><K^0\ |-|\ \ol{K^0}><\ol{K^0}\ |$ by using the connection that field
theory establishes between the charge conjugate of a neutral meson and its
hermitian conjugate; it is shown to vanish,
with a restricted meaning to be precised in the text. This introduces an
ambiguity in the definition of the effective mass matrix, to which
arbitrary antisymmetric diagonal terms can be added (in the
$(K^0,\ol{K^0})$ basis).

The fourth section starts by general considerations about the
diagonalization of complex matrices. In particular, we show the the
diagonalization by a bi-unitary transformation (used for example for the
mass matrices of fermions), in which the $|\ in>$ eigenstates form an
orthogonal basis and the $<out\ |$ eigenstates another such basis,
 and the other procedure where $|\ in>$ and $<out\
|$ eigenstates form a bi-orthogonal basis are not equivalent ({\em i.e.}
they lead not only to different eigenstates but to different mass
eigenvalues).

Despite the lack of a definitive argument in favor of it, we make the choice
to work with a bi-orthogonal basis of eigenstates, like was done in
\cite{AlvarezGaumeKounnasLolaPavlopoulos}.

Then, we go to the neutral kaon system and use
the most general parametrization of the effective mass matrix
$M$ for $K_L$ and $K_S$ in the $(K^0,\ol{K^0})$ basis given in
\cite{AlvarezGaumeKounnasLolaPavlopoulos}, which entails that the
eigenstates form a bi-orthogonal basis.
We first deform $M$ into its $CPT$ invariant form (which in this framework
is always possible). We show that the sole condition of $T$ invariance (for
the starting mass matrix)
is enough for the eigenstates $K_1$ and $K_2$ of the deformed matrix to be
the $CP$ eigenstates $(K^0 + \ol{K^0})$ and $(K^0 - \ol{K^0})$ (the
relation between the two mass spectra is explicitly calculated):
if $T$ is broken, then the mass
eigenstates can never be ``deformed'' into $CP$ eigenstates; however, that
the eigenstates can decay into two {\em and} three pions does not exclude that
they are of a type which can be ``deformed'' back to $(K^0+\ol{K^0})$ and
$(K^0-\ol{K^0})$.
Accordingly, and since this definition appears to be free of
ambiguity, we argue in favor of defining the breaking of $CP$
invariance (in mixing) through the one of $T$.

Next, we study the most general type of allowed deformation of $M$,
irrespectively of $CPT$ conservation. The
condition that the new eigenstates match $CP$ eigenstates is shown to bring
back to the previous case of deformation to a $CPT$ invariant mass matrix.

The question whether they can be flavour eigenstates is then raised. It is
shown that this indeed can happen, but then the mass of the $K^0$ is no
longer
equal to that of its antiparticle, as could be expected from the allowed
violation of $CPT$.

In the fifth section, we perform a similar analysis starting from the (diagonal)
mass matrix the $(K_L, K_S)$ basis. We start by re-expressing the
``vanishing'' combination $|\ K^0><K^0\ |-|\ \ol{K^0}><\ol{K^0}\ |$ in
terms of $|\ K_L^{in}>$, $K_S^{in}>$, $<K_L^{out}\ |$ and $<K_S^{out}\ |$;
various deformations are then studied.

We first investigate whether the mass spectra of the starting and deformed
mass matrices can match. It is proved that, if one assumes $CPT$ invariance
at the start, masses and eigenstates match at the same time. However, if
$CPT$ invariance is not assumed, the masses can match but the eigenstates
are different.

Then, we ask whether the eigenstates of the deformed mass matrix can be
degenerate in mass. We show that, to any set of split $K_L$ and $K_S$
mesons can be associated another bi-orthonormal basis of degenerate
eigenstates, but which belong to two different well defined ``vanishing''
deformations of the starting Hamiltonian. Finally we particularize the study
to the case when $CPT$ invariance is assumed at the start.

After a brief conclusion emphasizing recent attempts to treat
neutral mesons in the framework of QFT, two appendices come back to the
following points:\l
- the relevance of mass matrices with degenerate eigenvalues in the treatment
of neutral kaon-like mesons, in particular to generate mass split $CP$
violating eigenstates; 
this appendix is an addendum to section \ref{section:qftham};\l
- neutral mesons  are usually treated with an over-complete basis of states;
  using the results of section three, some simple relations linking
$|\ K_L^{in}>, |\ K_S^{in}>, |\ K_L^{out}>, |\ K_S^{out}>$ and their
hermitian conjugates are established.

%%%%%%%%%%%%%%%%%%%%%%%%%%%%%%%%%%%%%%%%%%%%%%%%%%%%%%%%%%%%%%%%%%%%%%%%%%%%%%
\section{FROM QUANTUM FIELD THEORY TO EFFECTIVE HAMILTONIAN}
\label{section:qftham}
%%%%%%%%%%%%%%%%%%%%%%%%%%%%%%%%%%%%%%%%%%%%%%%%%%%%%%%%%%%%%%%%%%%%%%%%%%%%%%%

It cannot be questioned that, like other particles, all problems
pertaining to neutral kaons should be tackled within the framework of
quantum field theory (QFT). Nevertheless, in all textbooks and most papers,
including \cite{AlvarezGaumeKounnasLolaPavlopoulos}, the treatment is
performed in the formalism of quantum mechanics, working with an effective
Hamiltonian or even simply an effective mass matrix with dimension $[m]$.

In QFT, the only coherent way to define masses of particles is by the poles
of their full propagator \cite{JacobSachs}; the kinetic terms play naturally 
a major role in this definition. The Lagrangian having no special reason to
be diagonal in a given basis, the propagator of a
system of particles in interaction is, in general, a non diagonal matrix.
For the sake of convenience, one works in a basis in which the (bare) kinetic
terms are diagonal.
Let us call $L$ the inverse full propagator for $n$ scalar fields;
it is a $n \times n$ matrix that we write, in momentum space
\begin{equation}
L (p^2) = p^2{\mathbb I} - M^{(2)}(p^2),
\label{eq:L}
\end{equation}
where $\mathbb I$, which factorizes the kinetic terms, is the $n \times n$
 unit matrix,  and the mass
matrix $M^{(2)}(p^2)$ is the sum of a bare mass matrix (which does not depend
on $p^2$) and terms coming from higher order contributions (renormalized
self-energies), which introduce dependences on the momenta; in particular,
higher order corrections are likely to produce non-diagonal
kinetic terms, which we include in $M^{(2)}(p^2)$.
The superscript in $M^{(2)}$ has been
put into parenthesis to avoid confusion with an ordinary power: it 
indicates that all elements of $M^{(2)}$ have dimension $[mass]^2$.

$L(p^2)$ is diagonalized by an $n \times n$ matrix $V(p^2)$ according to
\begin{equation}
V^{-1}(p^2) L(p^2) V(p^2) = p^2 {\mathbb I} - M_d^{(2)}(p^2),
\end{equation}
where $M_d^{(2)}(p^2)$ is diagonal, $M_d^{(2)}(p^2)=diag\left(
m_1^{(2)}(p^2), m_2^{(2)}(p^2), \ldots, m_n^{(2)}(p^2)\right)$. The
$m_j^{(2)}(p^2)$ have dimension $[mass]^2$ and the ``$()$'' are optional in
their superscripts.

The $n$ functions  $m_{1,\ldots,n}^{(2)}(p^2)$
are obtained by solving the characteristic equation
$\det(M^{(2)}(p^2) -m^{(2)})=0$;
the $n$ corresponding eigenvectors $\psi_j(p^2)$ are then determined by
\begin{equation}
M^{(2)}(p^2) \psi_j(p^2) = m_j^{(2)}(p^2)\psi_j(p^2).
\label{eq:eq2M2}
\end{equation}
The $(mass)^2$ of the $n$ scalar fields are defined as the values of
$p^2$ satisfying the $n$ self-consistent equations
$\left((p^2 -m_1^{(2)}(p^2)=0),\ldots, (p^2- m_n^{(2)}(p^2)=0)\right)$;
we call then $\mu_1^2, \mu_2^2, \ldots , \mu_n^2$:
\begin{equation}
\mu_i^2 = m_i^{(2)}(p^2=\mu_i^2);
\label{eq:mui}
\end{equation}
they satisfy accordingly the equation 
\begin{equation}
\det\left(p^2{\mathbb I} -
  V^{-1}(p^2)M^{(2)}(p^2)V(p^2)\right)
= \det(p^2{\mathbb I} -M^{(2)}(p^2))
=\prod_j{\left(p^2 - m_j^{(2)}(p^2)\right)}=0;
\end{equation}
since the $p^2$ term in $L$ is proportional to the unit matrix,
the same matrix $V(p^2)$ diagonalizes both $L(p^2)$ and $M^{(2)}(p^2)$. 

The (independent of $p^2$) physical mass eigenstates $\varphi_i$ of $L(p^2)$
satisfy the equation
\begin{equation}
M^{(2)}(\mu_i^2)\varphi_i = \mu_i^2 \varphi_i.
\label{eq:M2eq}
\end{equation}
Matching (\ref{eq:M2eq}) and (\ref{eq:eq2M2}) entails
\begin{equation}
\varphi_i = \psi_i(p^2=\mu_i^2).
\label{eq:phipsi}
\end{equation}
However, once $p^2$ is fixed to a certain value $\mu_\alpha^2$,
each $M^{(2)}(\mu_\alpha^2)$ is a
constant $n\times n$ matrix which has itself in general $n$
distinct eigenvalues and eigenvectors
\footnote{unless all eigenvalues of $M^{(2)}(\mu_\alpha^2)$ are
degenerate to $\mu_\alpha^2$.\label{footnote:degen2}}
: $\varphi_{j1} \equiv \psi_1(\mu_j^2)$ corresponds to the eigenvalue
$\mu^{(2)}_{j1} = m_1^{(2)}(\mu_j^2)$, $\ldots$, $\varphi_{jn}
\equiv \psi_n(\mu_j^2)$ corresponds to the eigenvalue
$\mu^{(2)}_{jn} = m_n^{(2)}(\mu_j^2)$.

(\ref{eq:M2eq}) do not determine uniquely the 
physical eigenvectors: $\varphi_i$ is only one among the $n$ eigenvectors
of $M^{(2)}(\mu_i^2)$.
There is in particular no reason why the $(n-1)$ other eigenvalues and
$(n-1)$ other eigenvectors have anything to do with the system of particles
under concern.

{\bf Can an effective constant mass matrix be naturally and uniquely
introduced?}

Let us consider the matrix $M_d(p^2)$, the square root of
$M_d^{(2)}(p^2)$, and
\begin{equation}
M_f(p^2) = V(p^2) M_d(p^2) V^{-1}(p^2);
\end{equation}
$M_f(p^2)$ is also the square root of $M^{(2)}(p^2)$ since
\begin{equation}
M^{(2)}(p^2) = V(p^2) M_d^{(2)}(p^2) V^{-1}(p^2) = V(p^2) M_d(p^2) V^{-1}(p^2)
V(p^2) M_d(p^2) V^{-1}(p^2) = (M_f(p^2))^2.
\end{equation}
The diagonalization equation for $M_f$
\begin{equation}
M_f(p^2) \chi_j(p^2) = y_j\chi_j(p^2),
\end{equation}
entails, left-multiplying  by $M_f(p^2)$
\begin{equation}
(M_f(p^2))^2 \chi_j(p^2) = (y_j)^2\chi_j(p^2),
\end{equation}
which shows that the eigenvectors of $M_f(p^2)$ are the same as the ones of
$M^{(2)}(p^2)$: 
\begin{equation}
\chi_j(p^2) = \psi_j(p^2),
\end{equation}
and that its eigenvalues (which of course turn out to be functions of $p^2$)
 are the square roots of the ones of $M^{(2)}(p^2)$: 
\begin{equation}
(y_j(p^2))^2 = m_j^{(2)}(p^2).
\end{equation}
Like for $M^{(2)}$, once a scale  $p^2$ has been fixed to $\mu_j^2$,
the resulting constant matrix $M_f(\mu_j^2)$ has $n$ eigenvalues $\mu_{j1}
\equiv \sqrt{m_1^{(2)}(\mu_j^2)}$, $\ldots$, $\mu_{jn}=\sqrt{m_n^{(2)}(
\mu_j^2)}$ and $n$ corresponding eigenvectors
$\varphi_{j1} = \psi_1(\mu_j^2)$, $\ldots$, $\varphi_{jn} =
\psi_n(\mu_j^2)$ (the same as $M^{(2)}(\mu_j^2)$).

 The physical mass $\mu_j$ is only one among these
eigenvalues, and the corresponding physical particle is one among these
eigenvectors:
at the physical poles $p^2=\mu_j^2$, and naming like before
\begin{equation}
\varphi_j = \chi_j(p^2=\mu_j^2) \equiv \psi_j(p^2=\mu_j^2),
\end{equation}
the equation
\begin{equation}
M_f(\mu_j^2) \varphi_j = y_j(\mu_j^2)\varphi_j.
\label{eq:Mfeq}
\end{equation}
entails 
\begin{equation}
(M_f(\mu_j^2))^2\varphi_j \equiv M^{(2)}(\mu_j^2)\varphi_j
= (y_j(\mu_j^2))^2 \varphi_j.
\end{equation}
and, so,
\begin{equation}
\mu_j^2 = (y_j(\mu_j^2))^2.
\label{eq:mujyj}
\end{equation}
It is tempting to introduce  $M_f$ as an effective mass
matrix and it is the only one proposed in the literature \cite{Beuthe}.
This proposition is comforted by  (\ref{eq:mujyj}) and the result
 stressed in \cite{JacobSachs}, that the masses of particles
can be consistently
taken as the square roots of the poles of the full propagator
\footnote{ and this
includes  the case when these poles are complex, {\em i.e.}
the case of unstable particles; the 
equation $y_j = \sqrt{\mu_j^{(2)}}$  for complex $\mu_j^{(2)}$
determine not only the masses (real parts) but also the widths
(imaginary parts) of the unstable particles from the values of the
poles of their full propagators.}
.

But, as we have seen, for a system of $n$ interacting particles,
there is not a single $M_f$ at work but $n$ of them, the $M_f(p^2)$'s
evaluated at
the $n$  poles of the full propagator $p^2=\mu_j^2$ solving
the $n$ self-consistent equations (\ref{eq:mui}). So, there are $n$
possible choices for $M_{eff}$ (see footnote \ref{footnote:degen2})
 and there cannot be a one to one
correspondence between QFT and a constant single effective mass matrix.

It is often argued that, when dealing with nearly degenerate systems like
the neutral kaon system, everything should be regular and smooth.
This is forgetting that, precisely,
in the case of nearly degenerate eigenvalues, a small
variation of the mass matrix can have large consequences on its spectrum.
Very close matrices, with very close
eigenvalues, can have very different eigenvectors. We give below a very
elementary example of this.

Consider the simplified case of a system of two nearly degenerate
interacting mesons. Two possible effective $2\times 2$ matrices are at work,
$M_{f1}$ and $M_{f2}$.

Suppose that, for example in the $(K^0,\ol{K^0})$ basis
\begin{equation}
M_{f1} = \left(\begin{array}{cc} \rho -\epsilon & 0 \cr
                                       0    & \rho + \epsilon
         \end{array}\right),\quad
M_{f2} = \left(\begin{array}{cc}    \rho & \epsilon_1 \cr
                                  \epsilon_2  &  \rho
        \end{array}\right)
\end{equation}
with $\rho,\epsilon,\epsilon_1,\epsilon_2 \in {\mathbb R}$ and
$\epsilon, \epsilon_1,\epsilon_2 \ll \rho$.
The eigenvalues of $M_{f1}$ are $\rho \pm\epsilon$ and its eigenvectors
are $K^0$ and $\ol{K^0}$.
The eigenvalues of $M_{f2}$ are $\rho \pm\sqrt{\epsilon_1\epsilon_2}$ and
the ratio of the components of its eigenvectors are $\pm
\sqrt{\epsilon_1/\epsilon_2}$, which can be very different from the $0$ and
$1$ which occur for the eigenvectors of $M_{f1}$.

We conclude accordingly that, for $n$ nearly degenerate particles, even if
the $n$ possible effective matrices $M_{f1,\ldots,fm}$ are very close,
the $(n-1)$ spurious eigenstates of each of them have in general nothing to
do with the true physical eigenstates. Only one eigenvector per effective
$M_f$ matrix is a faithful description of one of the physical mass eigenstates.

{\bf Is it legitimate to consider an effective single $\mathbf{n\times n}$
mass matrix the spectrum and eigenstates of which involve enough
measurable parameters to match all possible masses and eigenstates of a
system of interacting particles together with their transformation
properties by discrete symmetries?}

The following discussion concerns more specifically the approach of 
\cite{AlvarezGaumeKounnasLolaPavlopoulos} which starts from a single
constant effective mass matrix for the system of neutral kaons, which
includes enough (measurable) parameters to fit all possibly detected
mass eigenvalues and eigenvectors.

For the sake of simplicity, we proceed with a system of only two
neutral interacting mesons.

From above, we have learned that two constant $2\times 2$ matrices are
involved: $M_{f1} \equiv M_f(\mu_1^2)$ and $M_{f2}\equiv M_f(\mu_2^2)$.
Each of them has in general two eigenvalues and two associated eigenstates;
one pair is physically relevant, and the other pair is ``spurious''. So,
$\mu_1$ and $\varphi_1$ are one among the two eigenvalues
$(\mu_{11},\mu_{12})$ and one among the two eigenvectors
$(\varphi_{11},\varphi_{12})$ of $M_{f1}$, $\mu_2$ and $\varphi_2$ are
one among the eigenvalues $(\mu_{21},\mu_{22})$ and one among the eigenvectors
$(\varphi_{21},\varphi_{22})$ of $M_{f2}$.

Discrete symmetries set constraints on the QFT Lagrangian, and, in
particular, on the full renormalized mass matrix $M^{(2)}(p^2)$;
they  have to be satisfied at all values of $p^2$, so in particular by the
two matrices $M^{(2)}(p^2=\mu_1^2)$ and $M^{(2)}(p^2=\mu_2^2)$,
or their ``square roots'' $M_{f1}$ and $M_{f2}$.
Accordingly, relations are expected  between the eigenvalues and/or
the eigenstates of $M_{f1}$, or between the ones of $M_{f2}$; they
are likely to share many similarities since they originate from general
constraints on the $q^2$-dependent mass matrix; 
but each of these relations connect one physical to one spurious quantity,
corresponding to a given value of $p^2$; no special constraint is
{\em a priori} expected between  two physical  parameters,
since they are linked to two different values of $p^2$ and
to two different constant $M_f$ mass matrices.

First, we make some remarks about $CP$ violation. Since both mass
eigenstates are detected to violate $CP$, $M_{f1}$ and $M_{f2}$ are
expected to be non-normal, and it should also be a property of $M_{eff}$ if
we introduce such a single constant effective mass matrix.

Suppose however that both $M_{f1}$ and $M_{f2}$ have a
degenerate spectrum of eigenvalues, still with $\mu_1^2 \not= \mu_2^2$,
and that
they have the simplest possible form, proportional to the unit matrix
\begin{equation}
M_{f1}=\left(\begin{array}{cc} \mu_1 & 0 \cr 0 & \mu_1\end{array}\right),
\quad
M_{f2}=\left(\begin{array}{cc} \mu_2 & 0 \cr 0 & \mu_2\end{array}\right);
\end{equation}
any linear combination of $K^0$ and $\ol{K^0}$, in particular $CP$ violating,
is a mass eigenstate for each of them;  still, both  are normal matrices.
Furthermore, nothing prevents $M^{(2)}(p^2)$ itself to be normal since
it is only required to have degenerate eigenvalues at two values of
$p^2$

\footnote{The fact that the eigenvalues at  given $p^2$'s are degenerate is
important since, otherwise, no normal matrix can have $CP$ violating
eigenstates. See also  appendix \ref{section:degemama}.}
.
At the same time, a description of two split physical $CP$ violating
eigenstates by a single constant effective mass matrix demands that
it be non normal.  The mismatch between the two points of view is patent.

We proceed with some remarks concerning $CPT$ and assume that
it is a true symmetry of nature.

Suppose  that we observe two mass split neutral mesons $K_1$ and ${K_2}$.
Let $K_1$ and ${S_1}$ be the non-degenerate  eigenstates of
$M_f(\mu_1^2)$ which respects the constraints of $CPT$ invariance,
and that the same occurs for  $K_2$, ${S_2}$ and $M_f(\mu_2^2)$;
${S_1}$ and $S_2$ are the ``spurious'' eigenstates,
respectively of $M_{f1}$ and $M_{f2}$.
Suppose furthermore that one discovers that $K_2 = \ol{K_1}$;
the  natural conclusion is that $CPT$ is violated in the $K_1-K_2$ system,
and any description by a single constant effective mass matrix should
account for that (see for example
\cite{AlvarezGaumeKounnasLolaPavlopoulos});
it is however unclear whether $M_{f1}$ and $M_{f2}$ should have to respect
the same criteria  since $S_1$ and $S_2$ are devoid of physical
significance.

So: {\em the problem of finding the mass spectrum and eigenstates of a
system of interacting scalars cannot in general be reduced to
the diagonalization of a single constant mass matrix.}

It is unfortunately illusory to believe that difficulties attached to the
effective Hamiltonian formalism can be avoided by working directly in QFT;
indeed, computations in quantum
field theory  are performed with fields (quarks) which are not
the particles (kaons); quark diagrams can be evaluated, but cannot be
used to determine the renormalized self-energies of neutral kaons without
interpolation and limiting procedures like PCAC, which introduce themselves
again other types of uncertainties
\footnote{See also the conclusion and references therein.}
.

We shall nevertheless work in the formalism of a single constant
 effective mass matrix, to show that there yet exist other ambiguities
attached to it; they are the main subject of this work.

%%%%%%%%%%%%%%%%%%%%%%%%%%%%%%%%%%%%%%%%%%%%%%%%%%%%%%%%%%%%%%%%%%%%%%%%%%%%%%
\section{HILBERT SPACE AND OPERATORS}
\label{section:spaces}
%%%%%%%%%%%%%%%%%%%%%%%%%%%%%%%%%%%%%%%%%%%%%%%%%%%%%%%%%%%%%%%%%%%%%%%%%%%%%%

\subsection{Generalities}
\label{subsection:generalities}
%%%%%%%%%%%%%%%%%%%%%%%%%%%%%%%%%%%%%%%%%%%%%%%%%%%%%%%%%%%%%%%%%%%%%%%%%%%%%

In the case of unstable particles, non-hermitian effective mass matrices
must be used (see for example \cite{Silva} and \cite{BrancoLavouraSilva}).
The experimental fact that $CP$ is broken entails furthermore that an
effective mass matrix for neutral cannot be normal; its right and left
eigenvectors accordingly differ. They are commonly called respectively
``in'' and ``out'' eigenstates.
In connexion with this, the formalism of ``bra'' $|\ >$ and ``ket'' $<\ |$ is
also commonly used.

The goal of this section is to establish a connection between the bra's and
ket's $|\ K^0>, |\ \ol{K^0}>, <K^0\ |, <\ol{K^0}\ |$ used for neutral kaons
in the effective hamiltonian formalism and
the field operators $\phi_{K^0}, \phi_{\ol{K^0}}, \phi_{K^0}^\dagger,
\phi_{\ol{K^0}}^\dagger$ for these same kaons which appear in a Lagrangian of
quantum field theory.

In a  QFT Hamiltonian $\cal H$, an (hermitian) mass term for degenerate (stable)
neutral kaons would write (up to an eventual constant coefficient)
\begin{equation}
{\cal H}_m(x) = m^2\left(\phi_{K^0}^\dagger(x) \phi_{K^0}(x) +
\phi_{\ol{K^0}}^\dagger(x) \phi_{\ol{K^0}}(x)\right).
\label{eq:HmQFT}
\end{equation}
The operator $\phi_{K^0}(x)$
destroys a $K^0$ at space-time point $x$, $\phi_{K^0}^\dagger(x)$ creates one,
and $\phi_{\ol{K^0}}(x)$ and $\phi_{\ol{K^0}}^\dagger(x)$ do the same for
 $\ol{K^0}$ (see (\ref{eq:expand}) below). These four
operators have dimension $[mass]$.

In an effective hamiltonian approach $H(t) = \int d^3\vec x {\cal H}(x)$ and in
the ``bra'' and ``ket'' formalism, an equivalent mass term writes
\begin{equation}
H_m = m (|\ K^0><K^0\ | + |\ \ol{K^0}><\ol{K^0}\ |).
\label{eq:Hm}
\end{equation}
As the effective $H_m$ has dimension $[mass]$, the dimension of the bra's
must be the inverse of the one of the ket's. One can for example take both
to be dimensionless.

The close similarity between the two expressions (\ref{eq:HmQFT}) and
(\ref{eq:Hm}) suggests correspondences between bra's, ket's, and field
operators.
QFT sets hermiticity and commutation relations between the latter;
we shall transpose them accordingly into equivalent relations between the
bra's and ket's.
They are at the origin of ambiguities in the definition of the effective
mass matrix, which consequently also concern its eigenvectors and its spectrum.

\subsection{Hermiticity and commutation relations}
\label{subsection:relations}
%%%%%%%%%%%%%%%%%%%%%%%%%%%%%%%%%%%%%%%%%%%%%%%%%%%%%%%%%%%%%%%%%%%%%%%%%%%%

The neutral kaon $|\ K^0 >$ and its charge conjugate partner
$|\ \ol{K^0}> \equiv {\it C}|\ K^0>$ are defined, within the framework of
the flavour $SU(3)$
symmetry, as pseudoscalar states with quantum numbers $(6+i7)$ and $(6-i7)$,
with strangeness respectively $(+1)$ and $(-1)$.

The states $|\ 6>$ and $|\ 7>$ are considered to form  an orthonormal basis of
a 2-dimensional, a priori complex, Hilbert space $\cal V$; the latter is
endowed with a natural binary product $<\ |\ >$ satisfying
\begin{eqnarray}
&&<6\ |\ 6> = 1 = <7\ |\ 7>,\cr
&&<6\ |\ 7> = 0 = <7\ |\ 6>,
\end{eqnarray}

which entail
\begin{eqnarray}
&&<K^0\ |\ \ol{K^0}> = 0 = <\ol{K^0}\ |\ K^0>,\cr
&&<K^0\ |\ {K^0}> = 1 = <\ol{K^0}\ |\ \ol{K^0}>.
\label{eq:norm1}
\end{eqnarray}
The metric matrix \cite{Lowdin} $\Delta$ has been chosen equal to $1$; hence,
the basis ${\bf B}\equiv\{|\ 6>,|\ 7>\}$ and the reciprocal basis
${\bf\tilde B}= {\bf B}\Delta^{-1}$ are identical: the Hilbert space
$\cal V$ and its dual are isomorphic.

Since, in particular, the binary product has to be hermitian, the dual of a
vector $|\ A>$ of $\cal V$ can be noted $<A\ |$ and is the
hermitian conjugate of $|\ A>$
\begin{equation}
<A\ |= |\ A>^\dagger;
\label{eq:dual}
\end{equation}
in particular for the basis vectors
\footnote{Remark: particles are defined in quantum mechanics as eigenstates
of mass and spin; they are the ones which should be used to build the Fock
space, and it was shown in \cite{GiuntiKimLee} that, in the case of mixing
when mass eigenstates become different from flavour eigenstates, the latter
are not adequate to define the Fock space. However, in the case of near
degeneracy, which is the case for neutral kaons, the problem becomes only
conceptual and has no sizable effect.}
\begin{equation}
<K^0\ | = |\ K^0>^\dagger,\quad <\ol{K^0}\ | = |\ \ol{K^0}>^\dagger.
\label{eq:dual1}
\end{equation}
At the level of operators,
the `bra's $<K^0\ |$ and $<\ol{K^0}\ |$ belong to the dual of the Hilbert
space spanned by the `ket's $|\ K^0>$ and $|\ \ol{K^0}>$ ({\em i.e.} the space
of linear functionals acting on the ket's); the ket's can also be considered
to belong to the dual of the dual, {\em i.e.} the set of functionals acting
on the bras; in this respect, they are themselves operators.
According to (\ref{eq:norm1}), $<\ol{K^0}\ |$ annihilates the ket $|\ K^0>$
to yield $0$; accordingly, we like to put it in a one-to-one correspondence
with the operator $\Phi_{K^0}$ which, too, destroys a
$|\ K^0>$ state; a similar correspondence is natural to establish between
$<K^0\ |$ and $\Phi_{\ol{K^0}}$, which both destroy a $|\ \ol{K^0}>$:
allowing for arbitrary phases $\delta$ and $\zeta$
\footnote{The remark at the end of this section
concerning the mass term for degenerate $K^0$ and $\ol{K^0}$ shows that the
normalization chosen here is adequate.}
\begin{equation}
<\ol{K^0}\ | \sim \frac{e^{i\delta}}{\varsigma}\Phi_{K^0},\quad
<{K^0}\ |\sim \frac{e^{i\zeta}}{\varsigma}\Phi_{\ol{K^0}},
\label{eq:cor1}
\end{equation}
where $\varsigma$ is an arbitrary (real) mass scale.

Because of (\ref{eq:dual1}), through hermitian conjugation,
one also gets a one-to-one correspondence between $|\ K^0>$ and
$(\Phi_{\ol{K^0}})^\dagger$, and between $|\ \ol{K^0}>$ and
$(\Phi_{K^0})^\dagger$:
\begin{equation}
|\ \ol{K^0}>\sim \frac{e^{-i\delta}}{\varsigma}(\Phi_{{K^0}})^\dagger,\quad
|\ {K^0}> \sim \frac{e^{-i\zeta}}{\varsigma}(\Phi_{\ol{K^0}})^\dagger.
\label{eq:cor2}
\end{equation}
The relation $<\ol{K^0}\ | \ol{K^0}>=1$ of (\ref{eq:norm1})
 is now in one to one correspondence
with $(\phi_{K^0}/\varsigma)(\phi_{K^0}^\dagger/\varsigma)=1$; acting with
this product of operators for example on the vacuum shows that the
correspondence that we have established is consistent and legitimate (the
same of course can be done with $\ol{K^0}$).

Now, the expansions of $\Phi_{K^0}$ and $\Phi_{\ol{K^0}}$ in terms of
creation $(a^\dagger,b^\dagger)$ and annihilation $(a,b)$ operators
satisfying the usual commutation relations
\begin{equation}
[a(\vec k),a^\dagger(\vec l)] = [b(\vec k),b^\dagger(\vec l)]
= (2\pi)^3 2k_0 \delta^3(\vec k - \vec l)
\label{eq:commute}
\end{equation}
write ($\gamma$ is an arbitrary phase)
\begin{eqnarray}
\Phi_{K^0(x)} &=& \int \frac{d^3 \vec k}{(2\pi)^3 2k_0}\left(
 a(\vec k)e^{-ik.x} + b^\dagger(\vec k)e^{ik.x}\right),\cr
\Phi_{\ol{K^0}(x)} &=& e^{-i\gamma}\int \frac{d^3 \vec l}{(2\pi)^3 2l_0}\left(
 b(\vec l)e^{-il.x} + a^\dagger(\vec l)e^{il.x}\right).
\label{eq:expand}
\end{eqnarray}
According to (\ref{eq:expand}) one can write
\begin{equation}
\Phi_{\ol{K^0}} = e^{-i\gamma} (\Phi_{K^0})^\dagger,
\label{eq:cconj}
\end{equation}
which entails, using (\ref{eq:cor1}),(\ref{eq:cor2}),
$e^{-i\zeta}<{K^0}\ | =e^{-i\gamma}(e^{-i\delta}<\ol{K^0}\ |)^\dagger$, or
\begin{equation}
<{K^0}\ |=   e^{i(\delta+\zeta-\gamma)} |\ \ol{K^0}>.
\label{eq:op1}
\end{equation}
By doing so, we have identified, as announced before, ${\cal V}$ and its dual.

(\ref{eq:cconj}) reflects the property that, up to a phase, ``creating'' a
$K^0$ is equivalent to ``destroying'' a $\ol{K^0}$, and {\em vice-versa}.

The next important step is that, using the commutation relations
(\ref{eq:commute}) of the
$a,b,a^\dagger,b^\dagger$ operators,
one finds from (\ref{eq:expand}) that the following commutator vanishes
\begin{equation}
[\Phi_{K^0},\Phi_{\ol{K^0}}]=0,
\end{equation}
which, using (\ref{eq:cconj}), can be rewritten
\begin{equation}
e^{-i\gamma}(\Phi_{\ol{K^0}})^\dagger \Phi_{\ol{K^0}}-
e^{-i\gamma}(\Phi_{K^0})^\dagger \Phi_{K^0}=0,
\end{equation}
or, using the correspondences (\ref{eq:cor1}), (\ref{eq:cor2}) and dropping
the overall factor $e^{-i\gamma}$ (the phases $\delta$ and $\zeta$ drop
out)
\begin{equation}
|\ {K^0}><{K^0}\ | -  |\ \ol{K^0}><\ol{K^0}\ | \sim 0.
\label{eq:null}
\end{equation}
The sign ``$\sim$'' in (\ref{eq:null}) must not be misinterpreted; it
cannot mean a strict identity since this would be in contradiction with the
closure relation $|\ \ol{K^0}><\ol{K^0}\ | +  |\ K^0><K^0\ | = 1$.
We shall give to it the restricted following meaning:\l
{\em Within the Hilbert space spanned by $K^0$ and $\ol{K^0}$,
effective Hamiltonians for the neutral kaon system can be freely deformed
by adding terms proportional to $|\ \ol{K^0}><\ol{K^0}\ | -  |\ K^0><K^0\ |$.}

Changing the effective mass matrix changes its eigenstates and, in
particular, modifies their transformation properties by the discrete
symmetries $C$ and $T$.
Such deformations will be investigated in the next sections.

In the common (and ambiguous) notation where
$K^0$ and $\ol{K^0}$ stand for the corresponding field
operators $\phi_{K^0}$ and $\phi_{\ol{K^0}}$, 
this is akin to saying that mass terms for neutral kaons are
defined up to an arbitrary factor $\rho(K^0\ol{K^0}-\ol{K^0} K^0)$
\footnote{Using the correspondences (\ref{eq:cor1},\ref{eq:cor2}) and
then (\ref{eq:cconj}), an effective
mass term for (degenerate) $K^0$ and $\ol{K^0}$ mesons
$m (|\ K^0><K^0\ | + |\ \ol{K^0}><\ol{K^0}\ |)$,
is in one-to-one correspondence, in QFT, with
$m\varsigma\left(e^{-i\zeta}(\phi_{\ol{K^0}})^\dagger e^{i\zeta}\phi_{\ol{K^0}}
 + e^{-i\delta}(\phi_{K^0})^\dagger e^{-i\delta}\phi_{K^0}\right)
= m\varsigma\left(e^{i\gamma}\phi_{K^0}\phi_{\ol{K^0}} +
e^{i\gamma}\phi_{\ol{K^0}}\phi_{K^0} \right)$. In the usual
notation  this rewrites as the hermitian combination (the phase $\gamma$
is needed according to (\ref{eq:cconj}) to make the mass term hermitian)
 $m\varsigma e^{i\gamma}(K^0\ol{K^0} + \ol{K^0}K^0)$, which entails
in particular that one must take $\varsigma = m$.
}
($\rho\in{\mathbb C}$ if the hermiticity of the Hamiltonian is no longer
required, like in the present work).

The ambiguity that has just been stressed  adds up the other ones
 attached to the system of neutral mesons:\l
- ambiguity linked to the effective hamiltonian formalism (section
  \ref{section:qftham});\l
- ambiguities associated to the different possible procedures for
  diagonalizing a complex mass matrix, which we shall overview in
subsection \ref{subsection:diago}.

%%%%%%%%%%%%%%%%%%%%%%%%%%%%%%%%%%%%%%%%%%%%%%%%%%%%%%%%%%%%%%%%%%%%%%%%%%%%%%
\section{THE MASS MATRIX IN THE $\bs{(K^0,\ol{K^0})}$ BASIS;
VARIOUS  DEFORMATIONS}
\label{section:M00deform}
%%%%%%%%%%%%%%%%%%%%%%%%%%%%%%%%%%%%%%%%%%%%%%%%%%%%%%%%%%%%%%%%%%%%%%%%%%%%%%

In most decays, it turns out that the observed mass eigenstates  are not $K^0$
and $\ol{K^0}$, but the so-called $K_L, K_S$; they are unstable,
with the consequence already mentioned that their effective  mass matrix
$M$ has to be chosen non-hermitian \cite{BrancoLavouraSilva}, and
furthermore non-normal if $CP$ is violated..
The right eigenstates (or ``$|\ in>$'' states) and the left eigenstates
(or ``$<out\ |$'' states) -- their eigenvalues being the same -- are defined by
\begin{equation}
M |\ in> = \lambda |\ in>, \quad <out\ | M = <out\ | \lambda. 
\label{eq:inout}
\end{equation}
Though different, the $|\ in>$ and $|\ out> \equiv <out\ |
^\dagger$ states can be expanded on the same orthonormal basis formed by
$|\ K^0>$ and $|\ \ol{K^0}>$ (we already stressed that
the two Hilbert spaces ${\cal V}_{in}$ and
its dual ${\cal V}_{out}$ can be identified).

\subsection{Diagonalization of a complex matrix}
\label{subsection:diago}
%%%%%%%%%%%%%%%%%%%%%%%%%%%%%%%%%%%%%%%%%%%%%%%%%%%%%%%%%%%%%%%%%%%%%%%%%%%%%%

In the present context, it is relevant to briefly recall
the different ways of diagonalizing a complex mass matrix, and to
show that they correspond to different spectra and different eigenstates.

\subsubsection{Bi-unitary transformation}
\label{subsubsection:biunitary}
%%%%%%%%%%%%%%%%%%%%%%%%%%%%%%%%%%%%%%%%%%%%%%%%%%%%%%%%%%%%%%%%%%%%%%%%%%%%

Any complex mass matrix $M$ can be diagonalized by a bi-unitary
transformation:\l
$\forall M \in {\mathbb C}, \exists U\ and\  \exists V, U^\dagger M V = D,
UU^\dagger=1=VV^\dagger$
\footnote{$U$ and $V$ can be chosen to diagonalize respectively
$MM^\dagger$ and $M^\dagger M$.}
.

$U$ and $V$ are not unique, and can in particular be adapted such that the
elements of the diagonal matrix $D$ are all real. Indeed:\l
the {\em polar decomposition theorem} \cite{Lowdin}
 states that any complex matrix $M$ can
be written as the product $M=H\Lambda$, with $H$ hermitian
$H=H^\dagger$ and $\Lambda$ unitary $\Lambda\Lambda^\dagger=1=
\Lambda^\dagger\Lambda$.
Take $V=\Lambda^\dagger U$; one has:
$U^\dagger M V = U^\dagger H\Lambda\Lambda^\dagger U = U^\dagger H U=D$;
since $H$ is hermitian, its eigenvalues are real and $D$ is real;
$U$ is accordingly obtained by diagonalizing $H$.

Since $U$ and $V$ are unitary, the right and left eigenstates of $M$
 can be tuned to form two separate orthogonal sets; the 
counterpart of this is that the $|\ in>$ and $<out\ |$ eigenstates are not
orthogonal $<out\ |\ in>\not=0$,
 and that their bilinear products evolve with time.

Bi-unitary transformations are specially useful to diagonalize mass
matrices of fermions;
$U$ acts in this case on left-handed fermions and $V$ on right-handed ones (the
kinetic terms for the two chiralities are distinct).
In general, the fermion masses so defined ({\em i.e.}
the elements of the diagonal matrix obtained by  bi-unitary transformations)
do not coincide with the eigenvalues of the mass matrix
(see for example \cite{MachetPetcov}). They however do match
in the standard electroweak model because arbitrary rotations can always be
performed on right-handed quarks, which can in particular absorb the
unitary matrix $\Lambda$ occurring in the polar decomposition theorem (see
above) and bring back the mass matrix to a hermitian one.

\subsubsection{Bi-orthogonal basis}
\label{subsubsection:biorthogonal}
%%%%%%%%%%%%%%%%%%%%%%%%%%%%%%%%%%%%%%%%%%%%%%%%%%%%%%%%%%%%%%%%%%%%%%%%%%%%

The process of diagonalization that is used in
\cite{AlvarezGaumeKounnasLolaPavlopoulos} and in the present work
(see subsection \ref{subsection:biortho} below) is
not equivalent to a bi-unitary transformation
\footnote{In this respect, we disagree here with the footnote 1 on page 1 of
\cite{AlvarezGaumeKounnasLolaPavlopoulos}; see subsection
\ref{subsubsection:connection}.\label{footnote:foot1AG}}
. 

As can be seen on (\ref{eq:EVM}) below, neither the $U$ matrix linking $|\
K_L^{in}>$ and $|\ K_S^{in}>$ to $|\ K^0>$ and $|\ \ol{K^0}>$ nor $V$
linking $<K_L^{out}\ |$ and $<K_S^{out}\ |$ to $<K^0\ |$ and $<\ol{K^0}\ |$
is unitary; neither the $|\ in>$ nor the $<out\ |$ eigenstates form an
orthogonal set.
The orthogonality  relations involve {\em both} ``in'' and
``out'' eigenspaces, and they are now independent of time.

\subsubsection{Comparison between two procedures of diagonalization}
\label{subsubsection:connection}
%%%%%%%%%%%%%%%%%%%%%%%%%%%%%%%%%%%%%%%%%%%%%%%%%%%%%%%%%%%%%%%%%%%%%%%%%%%%

An extensive study of this question goes beyond the scope of the present
work and only a few points will be sketched out here.

The first unclear one is which diagonalization procedure has the cleanest
physical interpretation. Eventually which one should be used? We can
propose no specific answer here.

We shall rather insist of the fact that
not only the eigenstates associated to the two above procedures differ,
{\em but also their spectrum (eigenvalues)} -- see footnote
\ref{footnote:foot1AG} --.
We give below \cite{Thompson} some remarks concerning the links between
the two sets of eigenvalues in the $2\times 2$ case.

Let $M$ be a complex matrix, which can be written, according to the polar
decomposition theorem
\begin{equation}
M = H\Lambda,\quad with\  H=H^\dagger,
\Lambda\Lambda^\dagger=1=\Lambda^\dagger\Lambda,
\end{equation}
and let $\rho_1$, $\rho_2$ be the real eigenvalues of $M$ obtained by
a bi-unitary transformation and, in particular, by diagonalizing
the hermitian $H$ (see subsection \ref{subsubsection:biunitary})
\begin{equation}
U^\dagger H U = D \equiv \left(\begin{array}{cc} \rho_1 & \cr
                                            & \rho_2 \end{array}\right).
\end{equation}
On the other side, let $\lambda_1$ and $\lambda_2$ be the eigenvalues
of $M$ defined by the standard equations (\ref{eq:inout}) for $|\ in>$
eigenstates (which will be combined with $<out\ |$ eigenstates,
in the rest of the work, to form a bi-orthogonal basis)
\begin{equation}
M|\ \varphi_1^{in}> = \lambda_1 |\ \varphi_1^{in}>,\quad M|\ \varphi_2^{in}>
= \lambda_2 |\ \varphi_2^{in}>.
\end{equation}
$M$ rewrites
\begin{equation}
M= UD\tilde\Lambda U^\dagger \quad with\ \tilde\Lambda=U^\dagger\Lambda U,
\end{equation}
and we parametrize $\tilde\Lambda$ in the most general way
\begin{equation}
\tilde\Lambda= e^{i\theta}\left(\begin{array}{cc}
                   \cos \chi e^{i\phi} & \sin\chi e^{i\omega}\cr
                   -\sin\chi e^{-i\omega} & \cos\chi e^{-i\phi}
                   \end{array}\right).
\end{equation}
Evaluating the trace and the determinant of $M$ yields the two equations:
\begin{eqnarray}
&& Tr(M)\equiv \lambda_1+\lambda_2 =Tr(UD\tilde\Lambda U^\dagger)
= Tr(U^\dagger U D\tilde\Lambda)=Tr(D\tilde\Lambda) = e^{i\theta}\cos\chi
\left(\rho_1e^{i\phi}+\rho_2e^{-i\phi}\right), \cr
&& \det(M) \equiv \lambda_1\lambda_2 = \det(UD\tilde\Lambda U^\dagger)
=\det(D\tilde\Lambda) = \rho_1\rho_2 e^{2i\theta},
\end{eqnarray}
which yield the second order equation linking the $\lambda$'s and the
$\rho$'s
\begin{equation}
\lambda^2 -e^{i\theta}\cos\chi(\rho_1e^{i\phi}+\rho_2e^{-i\phi})
        +\rho_1\rho_2e^{2i\theta}=0.
\label{eq:lambdaeq}
\end{equation}
In the degenerate case
$\rho_1=\rho_2 =\rho$: one gets from (\ref{eq:lambdaeq})
\begin{equation}
\lambda_{deg} = e^{i\theta}\rho\left(
\cos\chi \cos\phi \pm i \sqrt{1-\cos^2\chi\cos^2\phi}\right)
=  e^{i(\theta\pm\alpha)}\rho\quad with\ \cos\alpha=\cos\chi\cos\phi,
\end{equation}
$\lambda_1$ and $\lambda_2$ are only deduced from $\rho_1$ and $\rho_2$
by multiplication by phases; this case is trivial and uninteresting.

To get somewhat further,
we simplify to the case when $H$ is already diagonal and its two
elements are very close $\rho_1=\rho=\rho_2-\epsilon$; one has then $U=1$,
$\tilde\Lambda\equiv U^\dagger\Lambda U = \Lambda$ and $V\equiv
\Lambda^\dagger U = \Lambda^\dagger$.

Developing the solutions of (\ref{eq:lambdaeq}) at first order in $\epsilon$,
one gets the following eigenvalues of $M$
\begin{eqnarray}
\lambda_\epsilon &=&e^{i\theta}\left[
\rho e^{\pm i\alpha}
+\frac{\epsilon}{2\sin\alpha}\left(\cos\phi\cos\chi\sin\alpha \mp
\sin\phi\cos\alpha \pm \sin\phi\right.\right. \cr
&&\hskip 2cm\left.\left. +i(-\sin\phi\cos\chi\sin\alpha
\mp\cos\phi\cos\alpha\pm\cos\phi)\right)\right]
\end{eqnarray}
The $\lambda_\epsilon$'s  are clearly not identical
to the elements of $D$ multiplied by phases; this proves that
diagonalization a complex matrix by a bi-unitary transformation
does not yield the same eigenvalues as looking for $|\ in>$ (and $<out\ |$)
eigenstates.

\subsection{Bi-orthogonal mass eigenstates for neutral kaons}
\label{subsection:biortho}
%%%%%%%%%%%%%%%%%%%%%%%%%%%%%%%%%%%%%%%%%%%%%%%%%%%%%%%%%%%%%%%%%%%%%%%%%%%%%

The mass matrix $M$ (in the $(K^0,\ol{K^0})$ basis)
\footnote{Both anti-diagonal terms were given the wrong sign in
the formula (22) of \cite{AlvarezGaumeKounnasLolaPavlopoulos}.
It has been corrected here: keeping the same expression for the mass
matrix, the eigenstates have been adjusted accordingly, taking also in
account our convention for charge conjugation of neutral kaons
$C|\ K^0> = + |\ \ol{K^0}>$; for example the state with $CP = -1$ is $(K^0 +
\ol{K^0})/\sqrt{2}$.}
\begin{equation}
M= \frac{1}{2} \left(\begin{array}{cc}
\Lambda - c & a + b \cr
 a - b &  \Lambda + c \end{array}\right),
\label{eq:M}
\end{equation}
in which the parameters $a$, $b$, $c$ and $\Lambda$ are given by
\begin{gather}
 a = (\lambda_L-\lambda_S)\frac{1+\alpha\beta}{1-\alpha\beta},\quad
 b = (\lambda_L-\lambda_S)\frac{\alpha+\beta}{1-\alpha\beta},\quad
 c = (\lambda_L-\lambda_S)\frac{\alpha-\beta}{1-\alpha\beta},\label{eq:elM1}\\
\Lambda = \lambda_L + \lambda_S,\label{eq:elM2}\\
\lambda_L \equiv \frac{\Lambda+\sqrt{a^2-b^2+c^2}}{2}
        = m_L -\frac{i\Gamma_L}{2},\quad
  \lambda_S \equiv \frac{\Lambda-\sqrt{a^2-b^2+c^2}}{2}
        = m_S -\frac{i\Gamma_L}{2},\label{eq:elM3}
\end{gather}
has the following right and left eigenvectors
 (``$|\ in>$'' and ``$<out\ |$'' states -- see (\ref{eq:inout}) --)

\vbox{
\begin{eqnarray}
|\ K_L^{in}> &=& \frac{1}{\sqrt{2(1-\alpha\beta)}}\left((1+\beta)|\ K^0> +(1-\beta)|\
\ol{K^0}>\right),\cr
|\ K_S^{in}> &=&  \frac{1}{\sqrt{2(1-\alpha\beta)}}\left((1+\alpha)|\ K^0> -(1-\alpha)|\
\ol{K^0}>\right),\cr
<K_L^{out}\ | &=&\frac{1}{\sqrt{2(1-\alpha\beta)}}\left((1-\alpha)<K^0\ | + (1+\alpha)<\ol{K^0}\
|\right),\cr
<K_S^{out}\ | &=&\frac{1}{\sqrt{2(1-\alpha\beta)}}\left((1-\beta)<K^0\ | - (1+\beta)<\ol{K^0}\
|\right),
\label{eq:EVM}
\end{eqnarray}
}
which are built to satisfy the orthogonality, normalization, and completeness
relations of a bi-orthonormal basis
\begin{eqnarray}
&&<K_L^{out}\ |\ K_L^{in}> = 1 = <K_S^{out}\ |\ K_S^{in}>,\cr
&&<K_L^{out}\ |\ K_S^{in}> = 0 = <K_S^{out}\ |\ K_L^{in}>,\cr
&& 1 = |\ K_L^{in}><K_L^{out}\ | + |\ K_S^{in}><K_S^{out}\ |.
\label{eq:norm}
\end{eqnarray}
The corresponding Hamiltonian is
\begin{equation}
{\cal H} = \lambda_L |\ K_L^{in}><K_L^{out}\ | + \lambda_S |\ K_S^{in}><K_S^{out}\ |.
\label{eq:H}
\end{equation}
$M$ becomes normal ($[M,M^\dagger]=0$) when $\beta =
-\alpha^\ast$\label{cond:normal}
\footnote{For any complex number $h$, $ h^\ast$ denotes its complex
conjugate.}
; in this
case the ``$|\ in>$'' and ``$|\ out>$'' eigenstates become identical.

(\ref{eq:EVM}) inverts into

\vbox{
\begin{eqnarray}
|\ K^0> &=& \frac{1}{\sqrt{2(1-\alpha\beta)}}
\left((1-\alpha)|\ K_L^{in}>+(1-\beta)|\ K_S^{in}>\right),\cr
|\ \ol{K^0}> &=&  \frac{1}{\sqrt{2(1-\alpha\beta)}}
\left((1+\alpha)|\ K_L^{in}> -(1+\beta)|\ K_S^{in}>\right),\cr
<K^0\ | &=&\frac{1}{\sqrt{2(1-\alpha\beta)}}
\left((1+\beta)<K_L^{out}\ | + (1+\alpha)<K_S^{out}\ |\right),\cr
<\ol{K^0}\ | &=&\frac{1}{\sqrt{2(1-\alpha\beta)}}
\left((1-\beta)<K_L^{out}\ | - (1-\alpha)<K_S^{out}\
|\right).
\label{eq:invEVM}
\end{eqnarray}
}
The contribution of (\ref{eq:M}) to the Hamiltonian (\ref{eq:H})
 proportional to $c$ writes $-(c/2)
(|\ K^0><K^0\ | - |\ \ol{K^0}><\ol{K^0}\ |)$; according to (\ref{eq:null}),
it can be dropped. More generally,
one may add to $M$ any such diagonal term, which has the net
effect of changing $c$ into $c-\epsilon$ and to transform $M$ given by
(\ref{eq:M}) into
\begin{equation}
N= \frac{1}{2}\left(\begin{array}{cc} \Lambda-(c-\epsilon) & a+b \cr
                          a-b & \Lambda+(c-\epsilon)\end{array}\right).
\label{eq:N}
\end{equation}
In the process of deforming $M$ into $N$, $\alpha$, $\beta$, $\lambda_L$ and
$\lambda_S$ are considered fixed, and, thus, $a$, $b$ and $c$ are fixed, too,
and are given in terms of the latter by (\ref{eq:elM1}).

The sum of the two eigenvalues is conserved in the deformation;
$\Lambda$ (see (\ref{eq:elM2})) is thus left invariant and $N$ can be
parametrized
in terms of new parameters $\tilde\alpha$, $\tilde\beta$, $\tilde a$,
$\tilde b$, $\tilde c$ by
\begin{equation}
N= \frac{1}{2}\left(\begin{array}{cc} \Lambda-\tilde c & \tilde a+\tilde b \cr
                          \tilde a-\tilde b & \Lambda+\tilde c \end{array}\right).
\label{eq:Ntilde}
\end{equation}
The ``tilde'' parameters are expressed in a way analogous to
(\ref{eq:elM1})
\begin{equation}
\tilde a = (\lambda_1-\lambda_2)\frac{1+\tilde\alpha\tilde\beta}
        {1-\tilde\alpha\tilde\beta},\quad
\tilde b = (\lambda_1-\lambda_2)\frac{\tilde\alpha+\tilde\beta}
        {1-\tilde\alpha\tilde\beta},\quad
\tilde c = (\lambda_1-\lambda_2)\frac{\tilde\alpha-\tilde\beta}
        {1-\tilde\alpha\tilde\beta},
\label{eq:abctilde}
\end{equation}
where $\lambda_1$ and $\lambda_2$ are the eigenvalues of $N$;
they  are expressed by
\begin{equation}
\lambda_1 = m_1 -\frac{i\Gamma_1}{2},\quad
  \lambda_2 = m_2 -\frac{i\Gamma_2}{2},\quad
\label{eq:elN}
\end{equation}
with $m_{1,2}$ (masses) and $\Gamma_{1,2}$ (widths) real,
and, since the sums of the diagonal terms of $N$ and $M$ are identical,
they satisfy
\begin{equation}
\lambda_1 + \lambda_2 = \Lambda = \lambda_L + \lambda_S.
\label{eq:delta}
\end{equation}
The eigenvectors $K_1,K_2$ of $N$ are accordingly parameterized in the
$(K^0,\ol{K^0})$ basis by

\vbox{
\begin{eqnarray}
|\ K_1^{in}> &=& \frac{1}{\sqrt{2(1-\tilde\alpha\tilde\beta)}}\left((1+\tilde\beta)|\ K^0> +
        (1-\tilde\beta)|\ \ol{K^0}>\right),\cr
|\ K_2^{in}> &=&  \frac{1}{\sqrt{2(1-\tilde\alpha\tilde\beta)}}\left((1+\tilde\alpha)|\ K^0> -
        (1-\tilde\alpha)|\ \ol{K^0}>\right),\cr
<K_1^{out}\ | &=&\frac{1}{\sqrt{2(1-\tilde\alpha\tilde\beta)}}\left((1-\tilde\alpha)<K^0\ | +
        (1+\tilde\alpha)<\ol{K^0}\ |\right),\cr
<K_2^{out}\ | &=&\frac{1}{\sqrt{2(1-\tilde\alpha\tilde\beta)}}\left((1-\tilde\beta)<K^0\ | -
        (1+\tilde\beta)<\ol{K^0}\ |\right),
\label{eq:EVN}
\end{eqnarray}
}
and the orthogonality and completeness relations satisfied by the eigenstates
(\ref{eq:EVN}) of the deformed matrix $N$ are, by construction, the same as
the ones of $M$
\begin{eqnarray}
&&<K_1^{out}\ |\ K_1^{in}> = 1 = <K_2^{out}\ |\ K_2^{in}>,\cr
&&<K_1^{out}\ |\ K_2^{in}> = 0 = <K_2^{out}\ |\ K_1^{in}>,\cr
&& 1 = |\ K_1^{in}><K_1^{out}\ | + |\ K_2^{in}><K_2^{out}\ |.
\label{eq:norm2}
\end{eqnarray}
The corresponding Hamiltonian is
\begin{equation}
\tilde{\cal H} =
\lambda_1 |\ K_1^{in}><K_1^{out}\ | + \lambda_2 |\ K_2^{in}><K_2^{out}\ |.
\label{eq:Htilde}
\end{equation}
From the two equivalent expressions (\ref{eq:N}) and (\ref{eq:Ntilde})
  of  $N$, one gets two expressions for
its eigenvalues $\lambda_1,\lambda_2$, respectively in terms of the
``untilde'' and ``tilde'' parameters
\begin{equation}
\lambda_1, \lambda_2=\frac{\Lambda \pm \sqrt{\tilde a^2-\tilde b^2+\tilde c^2}}{2}
          =  \frac{\Lambda \pm \sqrt{a^2 - b^2 + (c-\epsilon)^2}}{2},
\label{eq:lambdaCPT}
\end{equation}
and, so
\begin{equation}
\lambda_1-\lambda_2 = \sqrt{\tilde a^2 -\tilde b^2 + \tilde c^2}
              =  \sqrt{a^2 - b^2 + (c-\epsilon)^2}.
\label{eq:deltalambda}
\end{equation}
This ``invariance'' of the mass matrix by the above deformation
has the following consequences:\l
- the mass matrix of the neutral kaon system can always be transformed such
  that the criteria of $CPT$ invariance are fulfilled
(its diagonal terms can be made identical in the $(K^0,\ol{K^0})$ basis);\l
- there does not exist a one-to-one correspondence between mass matrix
  and observed mass eigenstates: indeed, deforming the mass matrix
changes the eigenstates, and the ones (\ref{eq:EVN}) of $N$ are not the
ones of $M$ given by (\ref{eq:EVM}).

%Let us examine them separately.

\subsection{Deformation of $\bs M$ into a  mass matrix
fulfilling the criteria of $\bs{CPT}$ invariance}
\label{subsection:CPT}
%%%%%%%%%%%%%%%%%%%%%%%%%%%%%%%%%%%%%%%%%%%%%%%%%%%%%%%%%%%%%%%%%%%%%%%%%%%%%%

The mass matrix $N_{CPT}$ below is obtained from $M$ by taking $\epsilon=c$
in (\ref{eq:N}):
\begin{equation}
N_{CPT}= \frac{1}{2} \left(\begin{array}{cc}
       \Lambda  & a + b \cr
    a - b & \Lambda        \end{array}\right);
\label{eq:NCPT}
\end{equation}
$a$ and $b$ are given by (\ref{eq:elM1}) in terms of $\alpha$, $\beta$,
$\lambda_L$ and $\lambda_S$ which are considered to be fixed.
The eigenvalues of $N_{CPT}$ (\ref{eq:NCPT}) are
\begin{equation}
\lambda_1^{CPT} = \frac{\Lambda + \sqrt{a^2-b^2}}{2},\quad
\lambda_2^{CPT}=\frac {\Lambda - \sqrt{a^2-b^2}}{2};
\label{eq:lambda12}
\end{equation}
their knowledge enables to calculate its eigenvectors,
which are parametrized according to (\ref{eq:EVN}) by $\tilde\alpha$ and
$\tilde\beta$.
One finds explicitly
\begin{equation}
\tilde\alpha_{CPT} = \tilde\beta_{CPT} =
        \frac{\sqrt{a+b}-\sqrt{a-b}}{\sqrt{a+b}+\sqrt{a-b}}
 \overset{(\lambda_L-\lambda_S)\not=0}{=}
\frac{\alpha+\beta}{1+\alpha\beta+\sqrt{1+\alpha^2\beta^2-\alpha^2-\beta^2}},
\label{eq:alphabetatildeCPT}
\end{equation}
and, for the normalization factor $\tilde n$ (see (\ref{eq:EVN}))
\begin{equation}
\tilde n_{CPT} \equiv \sqrt{2(1-\tilde\alpha_{CPT}\tilde\beta_{CPT})} =
\sqrt{2(1-\tilde\alpha_{CPT}^2)}= \frac{2\sqrt{2(a^2-b^2)}}{\sqrt{a+b} +
\sqrt{a-b}}.
\label{eq:ntildeCPT}
\end{equation}
That $\tilde\alpha_{CPT} = \tilde\beta_{CPT}$ entails, through (\ref{eq:abctilde}),
 that $\tilde c$ occurring in
(\ref{eq:Ntilde}) vanishes, $\tilde c = 0$, such that $N_{CPT}$ can finally
be parametrized by (see (\ref{eq:Ntilde}))
\begin{equation}
N_{CPT}= \frac{1}{2} \left(\begin{array}{cc}
                   \Lambda  & \tilde a_{CPT} +\tilde b_{CPT} \cr
  \tilde a_{CPT} - \tilde b_{CPT} & \Lambda      \end{array}\right),
\label{eq:NCPTtilde}
\end{equation}
which yields $\tilde a_{CPT} = a$ and $\tilde b_{CPT} = b$ by comparison with
(\ref{eq:NCPT}).

Its eigenvalues (\ref{eq:lambda12}) are then also given, from the
characteristic equation of (\ref{eq:NCPTtilde}), by
\begin{equation}
\lambda_1^{CPT} = \frac{\Lambda + \sqrt{\tilde a_{CPT}^2-\tilde b_{CPT}^2}}{2},\quad
\lambda_2^{CPT}=\frac {\Lambda - \sqrt{\tilde a_{CPT}^2-\tilde b_{CPT}^2}}{2}.
\label{eq:lambda12bis}
\end{equation}
One has in particular
\begin{equation}
\tilde a_{CPT}^2 - \tilde b_{CPT}^2 = a^2 - b^2,
\label{eq:abab}
\end{equation}
and we evaluate both sides of (\ref{eq:abab}).
Since $\tilde\alpha_{CPT} = \tilde\beta_{CPT}$ (see (\ref{eq:alphabetatilde})),
$\tilde a_{CPT}$ and $\tilde b_{CPT}$ are given, according to (\ref{eq:abctilde}), by
\begin{equation}
\tilde a_{CPT} = (\lambda_1^{CPT}-\lambda_2^{CPT}) \frac{1+\tilde\alpha_{CPT}^2}{1-\tilde\alpha_{CPT}^2},\quad
\tilde b_{CPT} = (\lambda_1^{CPT}-\lambda_2^{CPT})\frac{2\tilde\alpha_{CPT}}{1-\tilde\alpha_{CPT}^2},\quad
\label{eq:abtilde}
\end{equation}
such that $\tilde a_{CPT}^2 - \tilde b_{CPT}^2 =
(\lambda_1^{CPT}-\lambda_2^{CPT})^2$.
As far as the r.h.s. of (\ref{eq:abab}) is concerned, (\ref{eq:elM1})
entails that
$(a^2 - b^2) = (\lambda_L-\lambda_S)^2(1+\alpha^2\beta^2-\alpha^2-\beta^2)/
(1-\alpha\beta)^2$.

Finally, (\ref{eq:abab}) yields the ratio of the splittings between the
eigenvalues of the ``$CPT$ invariant'' and ``$CPT$ non-invariant'' mass matrices
$N_{CPT}$ and $M$
\begin{equation}
\frac{\lambda_1^{CPT}-\lambda_2^{CPT}}{\lambda_L-\lambda_S} =
        \frac{(1+\alpha^2\beta^2-\alpha^2-\beta^2)^{\frac{1}{2}}}
        {1-\alpha\beta}.
\end{equation}
The two mass splittings vanish simultaneously; they become identical only for
$\alpha=\beta$, but this trivial case (except when $\alpha=\beta \rightarrow 1$,
see below) corresponds (see (\ref{eq:elM1})) to $c=0$,
{\em i.e.} starting from an already $CPT$ invariant mass matrix which does not
undergo any deformation.

We now answer the two questions:\l
- can the eigenstates of the deformed mass matrix be $CP$ eigenstates?\l
- can they be flavour eigenstates?

$\bullet$\ The condition for $K_1$ and $K_2$ to be $CP$ eigenstates (see
(\ref{eq:EVN})) is $\tilde\alpha_{CPT} =0$ and $\tilde\beta_{CPT}=0$.
(\ref{eq:alphabetatildeCPT}) then entails $b=0$, which can only occur\l
- either if $\lambda_L = \lambda_S$, but this first case corresponds to a
trivially diagonal $M$ ($a=b=c=0$);\l
- or if $\beta=-\alpha$ (see (\ref{eq:elM1}));
the masses of the eigenstates of the deformed mass matrix are then given by
(\ref{eq:lambdaCPT}):
$\lambda_1^{CP},\lambda_2^{CP} = (\lambda_L + \lambda_S \pm a)/2$.

This is summarized as follows: by deforming
their mass matrix into its equivalent $CPT$ invariant form, any $K_L$ and
$K_S$ mass eigenstates of the following form

\vbox{
\begin{eqnarray}
|\ K_L^{in}> &=& \frac{1}{\sqrt{2(1+\alpha^2)}}\left(|\ K^0+\ol{K^0}>-\alpha|\ K^0-\ol{K^0}>\right) \cr
|\ K_S^{in}> &=& \frac{1}{\sqrt{2(1+\alpha^2)}}\left(|\ K^0-\ol{K^0}>+\alpha|\ K^0+\ol{K^0}>\right) \cr
<K_L^{out}\ | &=&\frac{1}{\sqrt{2(1+\alpha^2)}}\left(<K^0+\ol{K^0}\ |-\alpha<K^0-\ol{K^0}\ |\right)\cr
<K_S^{out}\ | &=&\frac{1}{\sqrt{2(1+\alpha^2)}}\left(<K^0-\ol{K^0}\ |+\alpha<K^0+\ol{K^0}\ |\right),
\label{eq:EVMCP}
\end{eqnarray}
}
with masses $\lambda_L$ and $\lambda_S$, can be transformed into $CP$ invariant
eigenstates with masses
\begin{equation}
\lambda_1=\lambda_L + \alpha^2 \lambda_S,\quad
\lambda_2=\lambda_S -\alpha^2\lambda_L,\quad
\frac{\lambda_1^{CP}-\lambda_2^{CP}}{\lambda_L-\lambda_S}=
\frac{1-\alpha^2}{1+\alpha^2}
\end{equation}
(we have used the expression for $a$
corresponding to $\beta=-\alpha$ given by (\ref{eq:elM1}):
$a_{CP}= (\lambda_L -\lambda_S)(1-\alpha^2)/(1+\alpha^2)$).
Note that $(\alpha+\beta)=0$ is the condition for $T$
invariance outlined in \cite{AlvarezGaumeKounnasLolaPavlopoulos}; so, when
$T$ invariance is satisfied, the $CP$ eigenstates $(K^0 \pm \ol{K^0})$ are
always mass eigenstates.

Defining $CP$ violation in mixing (``indirect'' $CP$ violation) by the
property that the mass eigenstates are linear combinations of $CP$
eigenstates appears ambiguous;
in the present framework,  the decays $K_L \rightarrow 2\pi$ and/or
$K_S \rightarrow 3\pi$, which have always been linked with
$CP$ violation in mixing,  do not  provide a sufficient
characterization of the latter
\footnote{The states (\ref{eq:EVMCP}) are expected to decay into final
states of various $CP$; nevertheless, by a transformation which adds a
``vanishing'' contribution to the effective Lagrangian, the corresponding mass
matrix can be deformed into another one the eigenstates of which are the
$CP$ eigenstates $(K^0 \pm\ol{K^0})$.}
.

On the other side, the detection of $T$ violation has been proved
\footnote{in the framework of a single constant effective mass matrix}
\cite{AlvarezGaumeKounnasLolaPavlopoulos} by the CPLEAR collaboration
\cite{CPLEAR}; from what has been shown above, this proves that the mass
eigenstates can never be cast into $CP$ eigenstates, and thus characterizes
$CP$ violation in mixing
\footnote{However, no sign of $T$ violation has been observed in the decay
$K^+ \rightarrow \pi^0 \mu \nu$ \cite{KEK-E246}.}
.

Remark that, in addition to the normalization conditions (\ref{eq:norm}),
(\ref{eq:EVMCP}) entails that, if $\alpha$ is furthermore real
\begin{equation}
\alpha \in {\mathbb R} \Rightarrow \left\{
\begin{array}{c} <K_L^{in}\ |\ K_S^{in}>=1=<K_L^{in}\ |\ K_S^{in}>,\quad
        <K_S^{in}\ |\ K_L^{in}>=0,\cr
        <K_L^{out}\ |\ K_L^{out}>=1=<K_S^{out}\ |\ K_S^{out}>,\quad
                <K_S^{out}\ |\ K_L^{out}>=0;
\end{array}\right.
\end{equation}
this is in agreement with the general condition for $M$ to be normal
in (\ref{cond:normal}), which transforms into the condition of reality for
$\alpha$ in the case $\beta=-\alpha$ under
concern. Then, $|\ in>$ and $|\ out>$ eigenstates become identical.

$\bullet$\ $K_1$ and $K_2$ can match the flavour eigenstates $K^0$ and $\ol{K^0}$ if
$\tilde\alpha_{CPT} = \pm 1$ and $\tilde\beta_{CPT} = \pm 1$. The two cases
$\tilde\beta_{CPT}= -\tilde\alpha_{CPT}=\pm 1$ are excluded by
(\ref{eq:alphabetatildeCPT}) which requires $\tilde\alpha_{CPT} =
\tilde\beta_{CPT}$.

The case $\tilde\alpha_{CPT} = \pm 1 = \tilde\beta_{CPT}$ needs some remarks.
It corresponds, by (\ref{eq:alphabetatildeCPT}),
either to $a=b$ or to $a=-b$, that is, in both cases, to a triangular $M$.
The eigenvectors of $M$ are $K^0$ {\em or} $\ol{K^0}$ (and not $K^0$ and
$\ol{K^0}$), which corresponds to $\alpha$ {\em or} $\beta$ equal to $\pm 1$.
$N_{CPT}$ is a triangular matrix with its two eigenvalues (its two diagonal
entries) equal to $(\lambda_L + \lambda_S)/2$.
The two  points to be noticed are:\l
- both $|\ K_1^{in}>$ and $|\ K_2^{in}>$ are
proportional to $|\ K^0>$,  while both $<K_1^{out}\ |$ and $<K_2^{out}\ |$ are
proportional to $<\ol{K^0}\ |$;\l
- at first sight, $<K_1^{out}\ |\ K_1^{in}> = 0 = <K_2^{out}\ |\ K_2^{in}>$;
however, the explicit calculation shows that the normalization
$\tilde n_{CPT}$ (\ref{eq:ntildeCPT}) of the new eigenstates goes to $0$,
and the Hamiltonian (\ref{eq:Htilde}) is given, as expected, by
\begin{equation}
\begin{CD}
\tilde{\cal H}_{CPT}
%\overset{\tilde\alpha_{CPT}=\tilde\beta_{CPT}\rightarrow\pm 1}{\longrightarrow}
@>>{\tilde\alpha_{CPT}=\tilde\beta_{CPT}\rightarrow\pm 1}>
\end{CD}
 \frac{\lambda_L + \lambda_S}{2}\left(
|\ K^0><K^0\ | + |\ \ol{K^0}><\ol{K^0}\ |\right).
\end{equation}

\subsection{General deformation}
\label{subsection:general}
%%%%%%%%%%%%%%%%%%%%%%%%%%%%%%%%%%%%%%%%%%%%%%%%%%%%%%%%%%%%%%%%%%%%%%%%%%%%

We now study the more general case of the deformation of $M$ (\ref{eq:M})
into $N$ (\ref{eq:N}), for any $\epsilon$.

(\ref{eq:deltalambda}) provides, through (\ref{eq:elM1}),
 the relation -- in general non linear -- between the two mass splittings
$(\lambda_1-\lambda_2)$ (corresponding to the ``deformed'' mass matrix)
 and $(\lambda_L-\lambda_S)$ (corresponding to the original mass matrix),
and the parameters $\alpha$, $\beta$ and $\epsilon$
\begin{equation}
\lambda_1-\lambda_2 = \sqrt{(\lambda_L-\lambda_S)^2
-2\epsilon(\lambda_L-\lambda_S)\frac{\alpha-\beta}{1-\alpha\beta}+\epsilon^2}
\label{eq:delDel}
\end{equation}
For $(\lambda_L-\lambda_S)=0$, $(\lambda_1-\lambda_2)=\pm\epsilon$, but this trivial case
corresponds by (\ref{eq:elM1}) to $a=b=c=0$ and, hence, to $M$ diagonal.
For $(\lambda_L-\lambda_S) \not=0$, $(\lambda_1-\lambda_2)$ can vanish for
$\epsilon= (\lambda_L-\lambda_S) (\alpha-\beta
\pm\sqrt{\alpha^2+\beta^2-\alpha^2\beta^2-1})/(1-\alpha\beta)$.

After some algebra, the coefficients $\tilde\alpha$ and $\tilde\beta$
determining the new eigenvectors (see (\ref{eq:EVN})) are determined to be
\begin{equation}
\tilde\alpha=\frac{\sqrt{a^2-b^2+(c-\epsilon)^2} -(a-b)+(c-\epsilon)}
{\sqrt{2(a-b)(a+\sqrt{a^2-b^2+(c-\epsilon)^2}})},\quad
\tilde\beta=\frac{\sqrt{a^2-b^2+(c-\epsilon)^2} -(a-b)-(c-\epsilon)}
{\sqrt{2(a-b)(a+\sqrt{a^2-b^2+(c-\epsilon)^2}})},
\label{eq:alphabetatilde}
\end{equation}
which yields the normalization factor (see (\ref{eq:EVN}))
\begin{equation}
\tilde
n \equiv \sqrt{2(1-\tilde\alpha\tilde\beta)}
=2\sqrt{\frac{\sqrt{a^2-b^2+(c-\epsilon)^2}}{a+\sqrt{a^2-b^2+(c-\epsilon)^2}}}.
\end{equation}
(\ref{eq:alphabetatilde}) gives back, as expected,
(\ref{eq:alphabetatildeCPT}) when $\epsilon=c$.

From the expressions of $\tilde\alpha$ and $\tilde\beta$
(\ref{eq:alphabetatilde}), the ones of $\tilde a$, $\tilde b$ and $\tilde
c$ can be determined via (\ref{eq:abctilde}); this entirely determines the
matrix $N$ (\ref{eq:N})(\ref{eq:Ntilde}) in terms of $a$, $b$, $c$, and
$\epsilon$, or in terms of $\alpha$, $\beta$, $\lambda_L$, $\lambda_S$ and
$\epsilon$.

Thanks to the expressions (\ref{eq:alphabetatilde}), the question whether
one (or both) eigenstate(s) of the deformed mass matrix can be $CP$
eigenstate(s) or flavour eigenstate(s) can be answered. 

$\bullet$\ Like previously, the condition for $K_1$ and $K_2$ to be $CP$
eigenstates (see (\ref{eq:EVN})) is $\tilde\alpha_{CPT} =0$ and
$\tilde\beta_{CPT}=0$. This entails in particular
$\tilde\alpha-\tilde\beta=0$, which, by (\ref{eq:alphabetatilde}), can only
be achieved for $\epsilon=c$; this brings us back to subsection
\ref{subsection:CPT}.

$\bullet$\ $K_1$ and $K_2$ can match the flavour eigenstates $K^0$ and
$\ol{K^0}$ if $\tilde\alpha_{CPT} = \pm 1$ and $\tilde\beta_{CPT} = \pm
1$.\l
- The case $\tilde\beta=\tilde\alpha=\pm1$ requires $\epsilon=c$ for the same
reason as above that $\tilde\beta-\tilde\alpha$ has to vanish, and this
brings us back again to subsection \ref{subsection:CPT};\l
- the cases $\tilde\beta=-\tilde\alpha=\pm 1$ yield, using
(\ref{eq:alphabetatilde})
\begin{equation}
\left\{\begin{array}{c}
\tilde\alpha+\tilde\beta=0 \Rightarrow
                \sqrt{a^2-b^2+(c-\epsilon)^2}=a-b;\cr
\Rightarrow \big\{\tilde\alpha=-\tilde\beta=\pm 1 \Rightarrow \epsilon =
 c \pm \sqrt{2(a-b)(2a-b)}\big\}.
\end{array}\right.
\label{eq:alphabetaflagen}
\end{equation}
The two equations (\ref{eq:alphabetaflagen}) above can only be satisfied\l
- either if $a=b$, which entails $\epsilon=c$ and bring us back to the last
  case and subsection \ref{subsection:CPT};\l
- or if $a=0$, which entails $\epsilon = c \pm b\sqrt{2}$. $a=0$ can only be
  achieved either if $\lambda_L=\lambda_S$, or if $\beta=-1/\alpha$.
The case $\lambda_L=\lambda_S$ is uninteresting since it also corresponds
to $b=0=c$ and to a degenerate diagonal $M$. Restricting to $\beta=-1/\alpha$,
and using (\ref{eq:elM1}), one then gets
\begin{equation}
\epsilon =
(\lambda_L-\lambda_S)\frac{\alpha^2+1\pm \sqrt{2}(\alpha^2-1)}{2\alpha}.
\label{eq:epsilonflagen}
\end{equation}
In this last case, the ``flavour'' and $CP$ violating $K_L$ and $K_S$
eigenstates can both be defined as ``mass eigenstates''. It can be
summarized as follows:\l
by deforming their mass matrix from $M$ (\ref{eq:M}) to $N$ (\ref{eq:N})
with $\epsilon$ given by (\ref{eq:epsilonflagen})
, any $K_L$ and $K_S$ mass eigenstates of the form (see (\ref{eq:EVM}) with
$\beta=-1/\alpha$)

\vbox{
\begin{eqnarray}
|\ K_L^{in}> &=& \frac{1}{2}\left(|\ K^0+\ol{K^0}>-\frac{1}{\alpha}|\ K^0-\ol{K^0}>\right) \cr
|\ K_S^{in}> &=& \frac{1}{2}\left(|\ K^0-\ol{K^0}>+\alpha|\ K^0+\ol{K^0}>\right) \cr
<K_L^{out}\ | &=&\frac{1}{2}\left(<K^0+\ol{K^0}\ |-\alpha<K^0-\ol{K^0}\ |\right)\cr
<K_S^{out}\ | &=&\frac{1}{2}\left(<K^0-\ol{K^0}\ |+\frac{1}{\alpha}<K^0+\ol{K^0}\ |\right),
\label{eq:EVMflagen}
\end{eqnarray}
}
with masses $\lambda_L$ and $\lambda_S$, can be transformed into flavour
mass eigenstates $K^0$ and $\ol{K^0}$ with masses
\begin{eqnarray}
&&\lambda_1,\lambda_2 =\frac
{2\alpha(\lambda_L+\lambda_S)\pm(\alpha^2-1)(\lambda_L-\lambda_S)}{4\alpha},\cr
&&\lambda_1-\lambda_2 = \pm b,\quad
\frac{\lambda_1-\lambda_2}{\lambda_L-\lambda_S} = \pm
\frac{1 - \alpha^2}{2\alpha} = \mp \frac{1-\beta^2}{2\beta}
\end{eqnarray}
(we have used the r.h.s. of (\ref{eq:deltalambda})).

For $|\alpha|=1$, $M$ becomes normal since the condition
$\beta=-1/\alpha$ under concern then matches the one in
 (\ref{cond:normal}) and $|\ in>$ and $|\ out>$ states become identical.

$\lambda_1$ and $\lambda_2$ are in general non-degenerate (since they
correspond to the masses of a particle and its antiparticle, $CPT$ is
broken, as expected since the diagonal terms of the deformed mass matrix
$N$ are not equal for $\epsilon$ given by (\ref{eq:epsilonflagen}));
they become degenerate only for $\alpha=\pm 1$, which is the trivial case when
the starting $K_L$ and $K_S$ are themselves flavour eigenstates.

%%%%%%%%%%%%%%%%%%%%%%%%%%%%%%%%%%%%%%%%%%%%%%%%%%%%%%%%%%%%%%%%%%%%%%%%%%%%%%
\section{THE MASS MATRIX IN THE $\bs{(K_L,K_S)}$ BASIS;
VARIOUS DEFORMATIONS}
\label{section:MLSdeform}
%%%%%%%%%%%%%%%%%%%%%%%%%%%%%%%%%%%%%%%%%%%%%%%%%%%%%%%%%%%%%%%%%%%%%%%%%%%%%%

In this section, we start from the mass matrix in the $(K_L,K_S)$ basis,
which has the simplest possible form since it is, by definition, diagonal.
We deform it by adding contributions
proportional to (\ref{eq:null}), after rewriting it in terms of the
$K_L$ and $K_S$ $|\ in>$ and $<out\ |$ states.

We then investigate whether the eigenvalues of the deformed mass matrix can
match those of the starting one, and, last, whether the eigenvalues of the
deformed mass matrix can be degenerate.

\subsection{Back to an operatorial identity}
\label{subsection:operator}
%%%%%%%%%%%%%%%%%%%%%%%%%%%%%%%%%%%%%%%%%%%%%%%%%%%%%%%%%%%%%%%%%%%%%%%%%%%%%%
(\ref{eq:null}) rewrites in terms of $K_L$ and $K_S$ ``$|\ in>$'' and
``$<out\ |$'' states (use (\ref{eq:invEVM}))
\begin{eqnarray}
\frac{{\cal H}_0}{\rho}
\equiv&&(\beta-\alpha)\left(|\ K_L^{in}><K_L^{out}\ |-|\ K_S^{in}><K_S^{out}\
|\right)\cr
&&           +(1-\alpha^2)|\ K_L^{in}><K_S^{out}\ |
             + (1-\beta^2)|\ K_S^{in}><K_L^{out}\ |\cr
 \sim&&0,
\label{eq:null2}
\end{eqnarray}
where $\rho$ is an arbitrary mass parameter.

In the $(K_L,K_S)$ basis, by definition, whatever the parameters $\alpha,\beta$,
 the kaon mass matrix $M$ (\ref{eq:M}) becomes diagonal and writes
\begin{equation}
{\cal M}_0=
\left(\begin{array}{cc}    \lambda_L   &     0     \cr
                                0      &     \lambda_S \end{array}\right);
\label{eq:M0}
\end{equation}
according to (\ref{eq:null2}), it may be freely transformed into
\begin{equation}
{\cal M}(\rho)= {\cal M}_0
+ \rho 
\left(\begin{array}{cc}    \beta-\alpha   &     1-\alpha^2     \cr
                           1-\beta^2      &  \alpha-\beta \end{array}\right),
\label{eq:Mrho}
\end{equation}
the eigenvalues of which are
\begin{equation}
\mu_1,\mu_2= \frac
{\lambda_L+\lambda_S \pm
    \sqrt{(\lambda_L-\lambda_S)^2-4\rho\left((\lambda_L-\lambda_S)(\alpha-\beta)
           -\rho(1-\alpha\beta)^2\right)}}
{2}.
\label{eq:mu1mu2}
\end{equation}
One parametrizes ${\cal M}(\rho)$ by
\begin{equation}
{\cal M}(\rho)= \frac{1}{2} \left(\begin{array}{cc}
\Lambda - \underline c & \underline a + \underline b \cr
 \underline a - \underline b &  \Lambda + \underline c \end{array}\right),
\label{eq:Mpar}
\end{equation}
which defines unambiguously $\underline a$, $\underline b$ and $\underline
c$
\footnote{The sum of the diagonal terms is $\Lambda \equiv
(\lambda_L+\lambda_S)$ fixed. $\underline c$ is the difference of the
diagonal terms, $\underline a$ and $\underline b$ respectively the sum and
the difference of the anti-diagonal terms.}
.
For $\mu_2 \not=\mu_1$, one can introduce two parameters $\underline\alpha$ and
$\underline\beta$ such that
\footnote{The three parameters $\underline a$, $\underline b$ and
$\underline c$ are not independent since they satisfy
\begin{equation}
\underline a^2 -\underline b^2 +\underline c^2 = (\mu_2-\mu_1)^2.
\label{eq:abccons}
\end{equation}
They are exchanged for the two independent parameters $\underline\alpha$ and
$\underline \beta$ defined by (\ref{eq:elM1und}) like originally in the
$(K^0,\ol{K^0})$ basis for $M$.}
\begin{gather}
 \underline a = (\mu_2-\mu_1)\frac{1+\underline \alpha\underline \beta}{1-\underline \alpha\underline \beta},\quad
 \underline b = (\mu_2-\mu_1)\frac{\underline \alpha+\underline \beta}{1-\underline \alpha\underline \beta},\quad
 \underline c = (\mu_2-\mu_1)\frac{\underline \alpha-\underline \beta}{1-\underline \alpha\underline \beta}
\label{eq:elM1und}
\end{gather}
such that the eigenvectors of ${\cal M}(\rho)$ write 
\footnote{$|\ K_L^{in}>$, $|\ K_S^{in}>$, $<K_L^{out}\ |$ and $<K_S^{out}\
|$ have been built to form a bi-orthonormal basis.}

\vbox{
\begin{eqnarray}
|\ \underline K_1^{in}> &=& \frac{1}{\sqrt{2(1-\underline\alpha\underline\beta)}}
        \left((1+\underline\beta)|\ K_L^{in}> +(1-\underline\beta)|\ K_S^{in}>\right),\cr
|\ \underline K_2^{in}> &=& \frac{1}{\sqrt{2(1-\underline\alpha\underline\beta)}}
        \left((1+\underline\alpha)|\ K_L^{in}> -(1-\underline\alpha)|\ K_S^{in}>\right),\cr
<\underline K_1^{out}\ | &=&\frac{1}{\sqrt{2(1-\underline\alpha\underline\beta)}}
        \left((1-\underline\alpha)<K_L^{out}\ | + (1+\underline\alpha)<K_S^{out}\ |\right),\cr
<\underline K_2^{out}\ | &=&\frac{1}{\sqrt{2(1-\underline\alpha\underline\beta)}}
        \left((1-\underline\beta)<K_L^{out}\ | - (1+\underline\beta)<K_S^{out}\
|\right).
\label{eq:EVMunderline}
\end{eqnarray}
}
One defines the dimensionless parameters $\underline u$, $\underline v$ and
$\underline w$ by
\begin{equation}
\underline u\equiv\frac{\underline a}{\mu_1-\mu_2}
        =\frac{1+\underline \alpha\underline \beta}{1-\underline \alpha\underline \beta},\quad
\underline v\equiv\frac{\underline b}{\mu_1-\mu_2}
        =\frac{\underline \alpha+\underline \beta}{1-\underline \alpha\underline \beta},\quad
\underline w\equiv\frac{\underline c}{\mu_1-\mu_2}
        =\frac{\underline \alpha-\underline \beta}{1-\underline \alpha\underline \beta};
\label{eq:uvwund}
\end{equation}
they are constrained by (\ref{eq:abccons}) to live on the 3-dimensional sphere
\begin{equation}
\underline u^2 - \underline v^2 + \underline w^2=1.
\label{eq:uvwcons}
\end{equation}
$\underline\alpha$ and $\underline\beta$ are determined by
(\ref{eq:elM1und}) and write
\begin{equation}
\underline\alpha=\frac{\underline v+\underline w}{\underline u+1},\quad
\underline\beta=\frac{\underline v-\underline w}{\underline u+1},
\label{eq:alphabetaund}
\end{equation}
and satisfy also, because of (\ref{eq:uvwcons}) the equation
$\underline\alpha\underline\beta=(\underline u -1)/(\underline u +1)$.

For any $\underline u \not= \pm1$, $\underline\alpha$ and $\underline\beta$ can be parametrized by
\begin{eqnarray}
\underline\alpha &=& \sqrt{\frac{1-\underline u}{1+\underline u}}(\sinh{\eta}+\cosh{\eta}),\cr
\underline\beta &=& \sqrt{\frac{1-\underline u}{1+\underline u}}(\sinh{\eta}-\cosh{\eta}),
\label{eq:paramund}
\end{eqnarray}
where we have made the change of variables
\begin{equation}
\cosh{\eta} = \frac{\underline w}{\sqrt{1-\underline u^2}},\quad
\sinh{\eta}=\frac{\underline v}{\sqrt{1-\underline u^2}}.
\end{equation}
So, for any complex $\underline u\not=\pm1$ and $\eta$, all matrices
${\cal M}(\rho)$ (\ref{eq:Mpar}) which rewrite, using
(\ref{eq:elM1und}), (\ref{eq:uvwund}) and (\ref{eq:paramund}),
\begin{equation}
\underline{\cal M} = \frac{1}{2}\left(\begin{array}{cc}
 \lambda_L+\lambda_S -(\mu_1-\mu_2)\sqrt{1-\underline u^2}\cosh\eta &
 (\mu_1-\mu_2)(\underline u+\sqrt{1-\underline u^2}\sinh\eta)\cr
(\mu_1-\mu_2)(\underline u-\sqrt{1-\underline u^2}\sinh\eta) &
\lambda_L+\lambda_S +(\mu_1-\mu_2)\sqrt{1-\underline u^2}\cosh\eta
\end{array}\right)
\end{equation}
and which can be obtained by deformation of ${\cal M}_0$,
have the same eigenvalues $\mu_1$ and $\mu_2$ (\ref{eq:mu1mu2}), and their
eigenvectors are given by

\vbox{
\begin{eqnarray}
|\ \underline K_1^{in}>
        &=& \frac{1}{2}\left( \sqrt{1+\underline u}(|\ K_L^{in}>+|\ K_S^{in}>) +
        \sqrt{1-\underline u}(\sinh\eta-\cosh\eta)(|\ K_L^{in}> - |\ K_S^{in})\right),\cr
|\ \underline K_2^{in}>
        &=& \frac{1}{2}\left( \sqrt{1+\underline u}(|\ K_L^{in}>-|\ K_S^{in}>) +
        \sqrt{1-\underline u}(\sinh\eta+\cosh\eta)(|\ K_L^{in}> + |\ K_S^{in}>)\right),\cr
<\underline K_1^{out}\ |
        &=& \frac{1}{2}\left( \sqrt{1+\underline u}(<K_L^{out}\ |+<K_S^{out}\
        |)-\sqrt{1-\underline u}(\sinh\eta+\cosh\eta)(<K_L^{out}\
        |-<K_S^{out}\ |)\right),\cr
<\underline K_2^{out}\ |
        &=& \frac{1}{2}\left( \sqrt{1+\underline u}(<K_L^{out}\ |-<K_S^{out}\
        |)-\sqrt{1-\underline u}(\sinh\eta-\cosh\eta)(<K_L^{out}\
        |+<K_S^{out}\ |)\right).\cr
&&
\label{eq:EVMund}
\end{eqnarray}
}

\subsection{Matching the mass spectra}
\label{subsection:matching}
%%%%%%%%%%%%%%%%%%%%%%%%%%%%%%%%%%%%%%%%%%%%%%%%%%%%%%%%%%%%%%%%%%%%%%%%%%%%%%

Among all possible cases,  the ones when $\mu_1,\mu_2$
eventually match $\lambda_L$ and $\lambda_S$ deserve a special investigation.
This can occur for $\rho=\hat\rho$ with
\begin{equation}
\hat\rho = (\lambda_L-\lambda_S)\frac{\alpha-\beta}{(1-\alpha\beta)^2}
=\frac{c}{1-\alpha\beta},
\label{eq:rhohat}
\end{equation}
where we have used (\ref{eq:elM1}) for the last identity on the r.h.s..

Since $\hat\rho$ only vanishes ({\em i.e.} the two mass spectra match and
there is no deformation)  when $\beta=\alpha$,
{\em i.e.} when the starting mass
matrix respects $CPT$ invariance, any non-vanishing
deformation of a $CPT$ invariant mass matrix alters its mass spectrum.

The question arises of which eigenstates correspond in general
 to $\rho=\hat\rho$.
Parameterizing, as before, $\hat{\cal M}\equiv{\cal M}(\hat\rho)$ by
\begin{equation}
\hat{\cal M}= \frac{1}{2} \left(\begin{array}{cc}
\Lambda - \hat c & \hat a + \hat b \cr
 \hat a - \hat b &  \Lambda + \hat c \end{array}\right),
\label{eq:Mhat}
\end{equation}
equations (\ref{eq:elM1und}), (\ref{eq:uvwund}) and (\ref{eq:alphabetaund})
are changed into their equivalent with $\mu_1$ replaced by $\lambda_L$,
$\mu_2$ replaced by$\lambda_S$, and underlined parameters replaced
by ``hatted'' ones.

Matching (\ref{eq:Mrho}) for $\rho=\hat\rho$ with (\ref{eq:Mhat}) yields
the equations linking  $\alpha$ and $\beta$ (the coefficients of
the eigenvectors (\ref{eq:EVM}) of $M$, in the $(K^0,\ol{K^0})$ basis),
to $\hat\alpha$
and $\hat\beta$ (the coefficients of the eigenvectors of the deformed
matrix in the $(K_L(\alpha,\beta),K_S(\alpha,\beta))$ basis).

\vbox{
\begin{eqnarray}
\frac{\hat\alpha-\hat\beta}{1-\hat\alpha\hat\beta} &=&
      2\left(\frac{\alpha-\beta}{1-\alpha\beta}\right)^2 -1,\cr
\frac{1+\hat\alpha\hat\beta+\hat\alpha+\hat\beta}{1-\hat\alpha\hat\beta}&=&
  2\frac{(\alpha-\beta)(1-\alpha^2)}{(1-\alpha\beta)^2},\cr
\frac{1+\hat\alpha\hat\beta-(\hat\alpha+\hat\beta)}{1-\hat\alpha\hat\beta}&=&
  2\frac{(\alpha-\beta)(1-\beta^2)}{(1-\alpha\beta)^2}.
\label{eq:abchat}
\end{eqnarray}
}
Calling
\begin{eqnarray}
\theta &=& 2\left(\frac{\alpha-\beta}{1-\alpha\beta}\right)^2 -1,\cr
\phi &=& (\alpha-\beta)\frac{2-\alpha^2-\beta^2}{(1-\alpha\beta)^2},\cr
\omega &=& (\alpha-\beta)\frac{\beta^2-\alpha^2}{(1-\alpha\beta)^2},
\label{eq:thetaphiomega}
\end{eqnarray}
$\theta$, $\phi$ and $\omega$ are of course non independent and satisfy the
constraint
\begin{equation}
\theta^2 + \phi^2 -\omega^2 -1 =0.
\label{eq:constpo}
\end{equation}
One finds explicitly
\begin{equation}
\hat\alpha=\frac{(\alpha-\beta) -1 + \beta^2}{(\alpha-\beta) +1 - \beta^2},
\quad
\hat\beta=\frac{-(\alpha-\beta) +1 - \alpha^2}{(\alpha-\beta) +1 -
\alpha^2}.
\end{equation}
As expected, for a $CPT$ invariant starting mass matrix $(\alpha=\beta)$,
the deformed matrix matches the starting one (we have seen above that
matching the mass spectrum requires $\hat\rho=0$), and their eigenstates
also match (one finds $\hat\alpha=-1, \hat\beta=1$).

For $\beta \not=\alpha$, the deformed mass matrix can keep the same mass
spectrum as the starting one, but exhibits different mass eigenstates.

\subsection{Degenerate ``deformed'' eigenstates}
\label{subsection:degenerate}
%%%%%%%%%%%%%%%%%%%%%%%%%%%%%%%%%%%%%%%%%%%%%%%%%%%%%%%%%%%%%%%%%%%%%%%%%%%%%

The values $\rho_d$ of $\rho$ which lead to degenerate mass eigenstates
(\ref{eq:mu1mu2}) for the deformed mass matrix, with mass
$\mu=(\lambda_L+\lambda_S)/2)$ are given by
\begin{equation}
\rho_d^2 -(\lambda_L-\lambda_S)\frac{\alpha-\beta}{(1-\alpha\beta)^2}\rho_d
    +\frac{1}{4}\left(\frac{\lambda_L-\lambda_S}{1-\alpha\beta}\right)^2=0.
\label{eq:rhodeq}
\end{equation}

\subsubsection{The general case}
\label{subsubsection:general}
%%%%%%%%%%%%%%%%%%%%%%%%%%%%%%%%%%%%%%%%%%%%%%%%%%%%%%%%%%%%%%%%%%%%%%%%%%%%%%

The solutions of (\ref{eq:rhodeq}) are
\begin{equation}
\rho_{d\pm}=(\lambda_L-\lambda_S)\frac{
(\alpha-\beta) \pm
\sqrt{(\alpha-\beta)^2-(1-\alpha\beta)^2}
}{2(1-\alpha\beta)^2}.
\end{equation}
Since $\mu_2=\mu_1$, the parametrization (\ref{eq:elM1und})
is no longer possible; the eigenstates of the deformed mass matrix have to
be investigated directly, in particular their normalization and orthogonality
which is no longer guaranteed.

The deformed mass matrix writes
\begin{eqnarray}
{\cal M}_\pm^{degen}&=& \frac{\lambda_L +\lambda_S}{2}\left(\begin{array}{cc}
                                                   1 & 0 \cr
                                                   0 & 1 \end{array}\right)
+ \frac{\lambda_L - \lambda_S}{2}\left(\begin{array}{cc}
        1-(\alpha-\beta)A_\pm   &    (1-\alpha^2)A_\pm        \cr
        (1-\beta^2)A_\pm        &  -(1-(\alpha-\beta)A_\pm) \end{array}\right)
\cr
&=& \frac{\lambda_L +\lambda_S}{2} D + \frac{\lambda_L -
\lambda_S}{2}\Delta
\label{eq:Mdegen}
\end{eqnarray}
with
\begin{equation}
A_\pm = \frac{(\alpha-\beta)\pm\sqrt{(\alpha-\beta)^2-(1-\alpha\beta)^2}}
{(1-\alpha\beta)^2}.
\label{eq:A}
\end{equation}
The determinant and the sum of the two diagonal terms of $\Delta$ vanishes,
ensuring that the eigenvalues of ${\cal M}_\pm^{degen}$ are
$(\lambda_L +\lambda_S)/2$.

Noting respectively the right and left eigenvectors
\footnote{To the matrix ${\cal M}_+^{degen}$ normally correspond two
right eigenvectors $|\ K_1^{in}>_+$ and $|\ K_2^{in}>_+$, and two left
eigenvectors $_+<K_1^{out}\ |$ and $_+<K_2^{out}\ |$; however, in the
present case, $|\ K_1^{in}>_+$ and $|\ K_2^{in}>_+$ are the same,
that we note $|\ K^{in}_+>$, and so is the case of $_+<K_1^{out}\ |$
and $_+<K_2^{out}\ |$, that we note $<K^{out}_+\ |$. The same occurs for
the eigenvectors of the matrix ${\cal M}_+^{degen}$, that we note $|\
K^{in}_->$ and $<K^{out}_-\ |$ \label{footnote:degen}.}
\begin{equation}
|\ K^{in}_\pm>= \frac{1}{n_\pm}\left(x_\pm |\ K_L^{in}> + y_\pm |\ K_S^{in}>\right),\quad
<K^{out}_\pm\ | = \frac{1}{\tilde n_\pm}\left(z_\pm <K_L^{out}\ | + t_\pm<K_S^{out}\ |\right)
\end{equation}
the coefficients $x,y,z,t$ must satisfy the conditions
\begin{eqnarray}
\frac{y_\pm}{x_\pm} &=&-\frac{1-(\alpha-\beta)A_\pm}{(1-\alpha^2)A_\pm}
        =\frac{(1-\beta^2)A_\pm}{1-(\alpha-\beta)A_\pm},\cr
\frac{t_\pm}{z_\pm}&=&-\frac{1-(\alpha-\beta)A_\pm}{(1-\beta^2)A_\pm}=
        \frac{(1-\alpha^2)A_\pm}{1-(\alpha-\beta)A_\pm},
\label{eq:condegen}
\end{eqnarray}
which immediately entail that the two following bilinear products vanish
\begin{equation}
<K^{out}_\pm\ |\ K^{in}_\pm> =0;
\label{eq:degenorm}
\end{equation}
since the ``in'' and ``out'' spaces have both shrunk to  1-dimensional
spaces (see footnote \ref{footnote:degen}), (\ref{eq:degenorm}) shows that
no suitable normalization and orthogonality condition can be achieved for
any single choice of $A$ ($A_+$ or $A_-$).

However, between eigenstates of different deformations of the starting mass
matrix, one gets
\begin{eqnarray}
<K^{out}_-\ |\ K^{in}_+> &=& 2\frac{x_+z_-}{n_+\tilde n_-},\cr
<K^{out}_+\ |\ K^{in}_-> &=& 2\frac{x_-z_+}{n_-\tilde n_+};
\end{eqnarray}
both bilinear products can be normalized to $1$ by the simplest choice
\begin{eqnarray}
%&& x_+=x_-=z_+=z_-=1,\cr
&& n_+=n_-=\tilde n_+=\tilde n_- = \sqrt{2},\cr
&& x_+z_-=1=x_-z_+;
\label{eq:norm3}
\end{eqnarray}
the last equation of (\ref{eq:norm3}) together with (\ref{eq:condegen})
entail
\begin{equation}
y_+t_-=1=y_-t_+.
\label{eq:norm4}
\end{equation}
A bi-orthonormal basis is formed by the four vectors $|\ K^{in}_+>$, $|\
K^{in}_->$, $<K^{out}_+\ |$ and $<K^{out}_-\ |$, corresponding to the same
mass but to two different deformations of the
starting mass matrix, respectively into ${\cal M}^{degen}_+$ and
${\cal M}^{degen}_-$ (\ref{eq:Mdegen}).

One gets  the closure relation
\begin{eqnarray}
|\ K^{in}_-><K^{out}_+\ | + |\ K^{in}_+><K^{out}_-\ |
&=& |\ K_L^{in}><K_L^{out}\ | + |\ K_S^{in}><K_S^{out}\ |\cr
&\equiv& |\ K^0><K^0\ | + |\ \ol{K^0}><\ol{K^0}\ |\cr
&=& 1,
\label{eq:closure2}
\end{eqnarray}
and
\begin{eqnarray}
&&|\ K^{in}_-><K^{out}_+\ | - |\ K^{in}_+><K^{out}_-\ |\cr
&&\cr
&& =
-\frac{\sqrt{(\alpha-\beta)^2-(1-\alpha\beta)^2}}{1-\beta^2}
|\ K_L^{in}><K_S^{out}\ |
-\frac{1-\beta^2}{\sqrt{(\alpha-\beta)^2-(1-\alpha\beta)^2}}
 |\ K_S^{in}><K_L^{out}\ |\cr
&&=
-\frac{1}{2(1-\alpha\beta)}\left[
\frac{\sqrt{(\alpha-\beta)^2-(1-\alpha\beta)^2}}{1-\beta^2}
\left((1-\beta)^2|\ \ol{K^0}><K^0\ |-(1+\beta)^2|\ K^0><\ol{K^0}\
|\right)\right.\cr
&&\left.\hskip 2cm+\frac{1-\beta^2}{\sqrt{(\alpha-\beta)^2-(1-\alpha\beta)^2}}
\left((1+\alpha^2)|\ K^0><\ol{K^0}\ | -(1-\alpha^2)|\ \ol{K^0}><K^0\ |\right)
\right];\cr
&&
\label{eq:diff}
\end{eqnarray}
in the last line of (\ref{eq:diff}) we have dropped the terms
proportional to $(|\ K^0><K^0\ |- |\ \ol{K^0}><\ol{K^0}\ |)$ according to
(\ref{eq:null}).
 
This subsection can be summarized as follows:\l
{\em To any set of split $K_L$ and $K_S$ mesons described by a Hamiltonian
$\cal H$ can be uniquely associated another bi-orthonormal basis of,
now degenerate, states which are the eigenstates of two different well
defined deformations of $\cal H$, ${\cal H} + \delta{\cal H}_\pm$; the
variations $\delta{\cal H}_\pm$ vanish
by the field theory constraints linking $K^0$ to $\ol{K^0}$.
There still exists a closure relation  which matches the one for
$K^0$ and $\ol{K^0}$, but which mixes the  two deformed Hamiltonians.}

\subsubsection{The case of a $\bs{CPT}$ invariant starting mass matrix}
\label{subsubsection:CPT}
%%%%%%%%%%%%%%%%%%%%%%%%%%%%%%%%%%%%%%%%%%%%%%%%%%%%%%%%%%%%%%%%%%%%%%%%%%%%%%

We study here the case $\alpha=\beta$, corresponding to a
$CPT$ conserving starting mass matrix $M$; one then gets
\begin{equation}
\rho_d^{CPT} = \pm\frac{i}{2}\frac{\lambda_L-\lambda_S}{1-\alpha^2},
\quad
A_\pm^{CPT}= \pm\frac{i}{1-\alpha^2}
\quad
\frac{y_\pm}{x_\pm}=\pm i,
\quad
\frac{t_\pm}{z_\pm}=\pm i.
\end{equation}
which corresponds, in the $(K_L,K_S)$ basis, to the deformed mass matrices
\begin{equation}
{\cal M}_{\pm\ CPT}^{degen} = \left(\begin{array}{cc}
      \lambda_L & \pm\frac{i}{2}(\lambda_L-\lambda_S) \cr
      \pm\frac{i}{2}(\lambda_L-\lambda_S) & \lambda_S \end{array}\right),
\end{equation}
with eigenvalues $\mu_1=\mu_2=(\lambda_L+\lambda_S)/2$.
The corresponding eigenvectors are
\footnote{(\ref{eq:diff}) rewrites:
$
%\begin{equation}
|\ K^{in}_-><K^{out}_+\ | - |\ K^{in}_+><K^{out}_-\ |=
i\left(|\ K_L^{in}><K_S^{out}\ | - |\ K_S^{in}><K_L^{out}\ |\right).
%\end{equation}
%
$
}
\begin{equation}
|\ K_\pm^{in}> = \frac{x_\pm}{\sqrt{2}}(|\ K_L^{in}> \pm i|\ K_S^{in}>),\quad
<K_\pm^{out}\ | = \frac{1}{\sqrt{2}}\frac{1}{x_\mp}
                          (<K_L^{out}\ | \pm i <K_S^{out}\ |),
\label{eq:degeig}
\end{equation}
or, writing $x_\pm = \sigma_\pm e^{i\chi_\pm}, \sigma_\pm \in
{\mathbb R}_+$
\begin{eqnarray}
|\ K_\pm^{in}> &=& \frac{\sigma_\pm}{\sqrt{2}}e^{i\chi_\pm}
                             (|\ K_L^{in}> \pm i|\ K_S^{in}>)\cr
&&\hskip 2cm= \frac{e^{i\chi_\pm}\sigma_\pm}{2\sqrt{1-\alpha^2}}
\left((1\pm i\alpha)|\ K^0+\ol{K^0}> + (\alpha\pm i)|\ K^0-\ol{K^0}>\right),\cr
<K_\pm^{out}\ | &=& \frac{1}{\sigma_\mp\sqrt{2}} e^{-i\chi_\mp}
                          (<K_L^{out}\ | \pm i <K_S^{out}\ |)\cr
&&\hskip 2cm=\frac{e^{-i\chi_\mp}}{2\sigma_\mp\sqrt{1-\alpha^2}}
\left((1 \mp i\alpha)<K^0+\ol{K^0}\ | +(-\alpha \pm i)<K^0-\ol{K^0}\
|\right).
\cr
&&
\label{eq:degeig2}
\end{eqnarray}
In the three cases $\{\alpha=0,\pm i\}$ below, either degenerate $CP$
eigenstates originate from split states of the form
$(K^0 \pm \ol{K^0}) \pm i (K^0\mp\ol{K^0})$
 , or split $CP$ eigenstates are
transformed into degenerate $(K^0 \pm \ol{K^0}) \pm i (K^0\mp\ol{K^0})$ states:
 
$\ast$\quad for $\alpha=0$, $K_L$ and $K_S$ are the $CP$ eigenstates
$K^0\pm\ol{K^0}$, and the degenerate states of (\ref{eq:degeig2}) become
\begin{eqnarray}
|\ K_\pm^{in}>_{\alpha=0} &=& \frac{e^{i\chi_\pm}\sigma_\pm}{2}
\left(|\ K^0+\ol{K^0}> \pm i|\ K^0-\ol{K^0}>\right),\cr
<K_\pm^{out}\ |_{\alpha=0} &=& \frac{e^{-i\chi_\mp}}{2\sigma_\mp}
\left(<K^0+\ol{K^0}\ |  \pm i<K^0-\ol{K^0}\ |\right);
\label{eq:degeig3}
\end{eqnarray}
$\ast$\quad for $\alpha=i$

\vbox{
\begin{eqnarray}
|\ K_L^{in}> &=& \frac{1}{2}
        \left(|\ K^0 + \ol{K^0}> +i\ |\ K^0 -\ol{K^0}>\right),\quad
|\ K_S^{in}> =  \frac{1}{2}
        \left(|\ K^0 - \ol{K^0}> +i\,|K^0 +\ \ol{K^0}>\right),\cr
<K_L^{out}\ | &=&\frac{1}{2}
        \left(<K^0+\ol{K^0}\ | -i\,<K^0-\ol{K^0}\ |\right),\quad
<K_S^{out}\ | =\frac{1}{2}
        \left(<K^0-\ol{K^0}\ | - i<K^0 +\ol{K^0}\ |\right),\cr
&&
\label{eq:EVM5}
\end{eqnarray}
}
and the degenerate eigenstates of the deformed mass matrices are

\vbox{
\begin{eqnarray}
|\ K_+^{in}>_{\alpha=i}&=&i\,\frac{\sigma_+ e^{i\chi_+}}{\sqrt{2}} |\ K^0-\ol{K^0}>,
\quad
|\ K_-^{in}>_{\alpha=i}=\frac{\sigma_- e^{i\chi_-} }{\sqrt{2}} |\ K^0+\ol{K^0}>,\cr
<K_+^{out}\ |_{\alpha=i} &=& \frac{e^{-i\chi_-}}{\sigma_-\sqrt{2}} <K^0+\ol{K^0}\ |,
\quad
<K_-^{out}\ |_{\alpha=i} = -i\,\frac{e^{-i\chi_+}}{\sigma_+\sqrt{2}} <K^0-\ol{K^0}\ |;
\label{eq:degeig4}
\end{eqnarray}
}
$\ast$\quad for $\alpha=-i$

\vbox{
\begin{eqnarray}
|\ K_L^{in}> &=& \frac{1}{2}
        \left(|\ K^0 + \ol{K^0}> -i\ |\ K^0 -\ol{K^0}>\right),\quad
|\ K_S^{in}> =  \frac{1}{2}
        \left(|\ K^0 - \ol{K^0}> -i\,|K^0 +\ \ol{K^0}>\right),\cr
<K_L^{out}\ | &=&\frac{1}{2}
        \left(<K^0+\ol{K^0}\ | + i\,<K^0-\ol{K^0}\ |\right),\quad
<K_S^{out}\ | =\frac{1}{2}
        \left(<K^0-\ol{K^0}\ | + i<K^0 +\ol{K^0}\ |\right),\cr
&&
\label{eq:EVM6}
\end{eqnarray}
}
and the degenerate eigenstates of the deformed mass matrices are

\vbox{
\begin{eqnarray}
|\ K_+^{in}>_{\alpha=-i}&=&\frac{\sigma_+ e^{i\chi_+}}{\sqrt{2}} |\ K^0+\ol{K^0}>,
\quad
|\ K_-^{in}>_{\alpha=-i}=-i\,\frac{\sigma_- e^{i\chi_-} }{\sqrt{2}} |\ K^0-\ol{K^0}>,\cr
<K_+^{out}\ |_{\alpha=-i} &=& i\,\frac{e^{-i\chi_-}}{\sigma_-\sqrt{2}} <K^0-\ol{K^0}\ |,
\quad
<K_-^{out}\ |_{\alpha=-i} = \frac{e^{-i\chi_+}}{\sigma_+\sqrt{2}} <K^0+\ol{K^0}\ |.
\label{eq:degeig5}
\end{eqnarray}
}

This example is the last that we propose to illustrate the many
aspects that the effective mass matrix for neutral kaons can take; of
course this study is not exhaustive and cannot be; 
a continuous range of different values is indeed possible, due to the relations
between the states of the basis which are provided
by the commutation relations of QFT,
and, in particular in the $(K_L,K_S)$ basis, because the corresponding
bi-orthogonal set of states is over-complete
(see appendix \ref{section:ocbasis}). It is only maybe useful to
emphasize again the dominant role of discrete symmetries to distinguish
between the different possible eigenstates.

%%%%%%%%%%%%%%%%%%%%%%%%%%%%%%%%%%%%%%%%%%%%%%%%%%%%%%%%%%%%%%%%%%%%%%%%%%%%%%%
\section{CONCLUSION}
\label{section:conclusion}
%%%%%%%%%%%%%%%%%%%%%%%%%%%%%%%%%%%%%%%%%%%%%%%%%%%%%%%%%%%%%%%%%%%%%%%%%%%%%%%

This work points at several  ambiguities occurring in the usual
treatment of  interacting neutral mesons.
After casting doubts on the validity of introducing a single constant
effective mass matrix, it was shown that, even if one puts aside the
corresponding arguments, new ambiguities arise which, in particular,
allow for deformations of the effective mass matrix;
one of the main consequences of them all is that the characterization
of $CP$ violation becomes itself blurred.

This adds to the longing for a fundamental explanation and theory of
$CP$ violation, which could go beyond the simple parameterization which we
have been standing on for many years.
It would be satisfactory  to exhibit, for example, a mechanism similar to
the phenomenon called ''frustration'' in solid state physics,
leading to an impossibility for mass eigenstates to align along
$CP$ conserving directions.

The ultimate goal is of course a treatment {\em ab initio} by QFT of the
systems of interacting mesons. The way to go is still long, but
beautiful attempts at such a program have appeared recently.  The works
\cite{BingerJi} \cite{BlasoneCapolupoRomeiVitiello}, which in particular
 uncover a non-trivial structure of the vacuum, specially deserve to be
mentioned, and it is to be hoped that, in a close future, they will be
improved and completed. We refer the reader to \cite{Beuthe} for their
analysis.

Last, I  mention that this work generalizes \cite{Machet}, which already
uncovered ambiguities  in the description of the neutral kaon system
by a constant effective mass matrix.

%%%%%%%%%%%%%%%%%%%%%%%%%%%%%%%%%%%%%%%%%%%%%%%%%%%%%%%%%%%%%%%%%%%%%%%%%%%%%%
\vskip .5cm

\centerline{\rule{7cm}{.2mm}}

\vskip .5cm 
\begin{em}
\underline {Acknowledgments}: it is a pleasure to thank G. Thompson for
enlightening discussions and advice, and L. Alvarez-Gaum\'e, who
in particular drew my attention to reference \cite{Beuthe}.
\end{em}

%%%%%%%%%%%%%%%%%%%%%%%%%%%%%%%%%%%%%%%%%%%%%%%%%%%%%%%%%%%%%%%%%%%%%%%%%%%%%%

%%%%%%%%%%%%%%%%%%%%%%%%%%%%%%%%%%%%%%%%%%%%%%%%%%%%%%%%%%%%%%%%%%%%%%%%%%%%%
\newpage\null
\appendix
%%%%%%%%%%%%%%%%%%%%%%%%%%%%%%%%%%%%%%%%%%%%%%%%%%%%%%%%%%%%%%%%%%%%%%%%%%%%%%

%%%%%%%%%%%%%%%%%%%%%%%%%%%%%%%%%%%%%%%%%%%%%%%%%%%%%%%%%%%%%%%%%%%%%%%%%%%%%%
\section{MASS MATRICES WITH DEGENERATE EIGENVALUES}
\label{section:degemama}
%%%%%%%%%%%%%%%%%%%%%%%%%%%%%%%%%%%%%%%%%%%%%%%%%%%%%%%%%%%%%%%%%%%%%%%%%%%%%%

This appendix refers more specifically to section \ref{section:qftham}.

$M^{(2)}(p^2)$ is the full renormalized mass matrix of the neutral kaon
system (see (\ref{eq:L})).
Suppose that it  degenerate eigenvalues {\em for all $p^2$},
{\em i.e.} that they satisfy
\begin{equation}
m^{(2)}_1(p^2) = m^{(2)}_2(p^2), \forall p^2.
\label{eq:degenp2}
\end{equation}

The  two physical masses $\mu_1^2$ and $\mu_2^2$ are defined by the two
equations
$m^{(2)}_1(p^2=\mu_1^2)=\mu_1^2, m^{(2)}_2(p^2=\mu_2^2)=\mu_2^2$,
which reduce to a single one with solution(s) $\mu^2$ when (\ref{eq:degenp2})
is satisfied;\l
- if its solution  $p^2=\mu^2$ is unique, there exists a single $M_f \equiv
  \left(M^{(2)}(\mu^2)\right)^{1/2}$ and the physical states are
always degenerate;\l
- if it has several solutions $p^2 = \mu_\alpha^2$, there exist as
  many  constant matrices $M_{f\alpha} \equiv M_f(p^2=\mu_\alpha^2)$
 and each of them has  degenerate eigenvalues;

If one writes, in the $(K^0,\ol{K^0})$ basis, such a generic $2\times 2$
constant mass matrix
%
%\begin{equation}
$M_{f\alpha} = \left(\begin{array}{cc}a_\alpha & b_\alpha
\cr c_\alpha & d_\alpha\end{array}\right)$, with $a,b,c,d \in {\mathbb C}$,
%\end{equation}
%
the conditions that its two eigenvalues coincide is
$(a_\alpha-d_\alpha)^2 + 4b_\alpha c_\alpha=0$. Let us take for example
$d_\alpha=a_\alpha-2i\sqrt{b_\alpha c_\alpha}$; the physical mass
is given by $\mu_\alpha = a_\alpha-i\sqrt{b_\alpha c_\alpha}$.

The right eigenvector
$ \left(\begin{array}{c}u_\alpha\cr v_\alpha\end{array}\right)
\equiv (u_\alpha |\ K^0> + v_\alpha|\ \ol{K^0}>)$ of $M_{f\alpha}$
 must satisfy $u_\alpha\sqrt{c_\alpha} =
iv_\alpha\sqrt{b_\alpha}$,
and its left eigenvector $(z_\alpha\ t_\alpha) \equiv (z_\alpha <K^0\ | + t_\alpha <\ol{K^0}\ |)$ must
satisfy $ z_\alpha\sqrt{b_\alpha} = it_\alpha\sqrt{c_\alpha}$.

{\bf Can all physical eigenstates be $\bs{CP}$ eigenstates?}

The only possible situation is that they appear in two different
\footnote{If they appear in more than two, certainly one $CP$ eigenstate will
occur  with two different masses, which is impossible. The eventual other
$M_{f\alpha}$ lead to spurious eigenvalues and eigenstates.}
constant effective matrices with degenerate eigenvalues
$M_{f1}$ and $M_{f2}$, and in such a way that the right eigenstate of
$M_{f1}$  do not match the one of $M_{f2}$, since we have supposed that
their masses are different $\mu_1 \not=\mu_2$.

The two above conditions on $u_\alpha, v_\alpha, z_\alpha, t_\alpha$
 entail that, if $u_\alpha/v_\alpha=1$, {\em i.e.}
if the right eigenstate is proportional to $|\ K^0 + \ol{K^0}>$, at the same
times one gets $z_\alpha/t_\alpha = -1$, such that the  left eigenstate is
proportional to $<K^0 - \ol{K^0}\ |$; this occurs for each $M_{f\alpha}$.
The corresponding picture is that
$|\ K^0 + \ol{K^0}>$ and $<K^0-\ol{K^0}\ |$ have mass $\mu_1$, and that
$|\ K^0 -\ol{K^0}>$ and $<K^0+\ol{K^0}\ |$ have mass $\mu_2$.

This situation is paradoxical since the two elements of  each
 pair of $|\ in>$ and $<out\ |$ eigenstates, which are supposed to represent
the same particle (with a given mass), have  exactly opposite $CP$ properties,
and are furthermore orthogonal, forbidding a suitable normalization and
closure relation. Rejecting this possibility, one concludes that:

{\em When the renormalized mass matrix for neutral kaons has 
degenerate ($p^2$ dependent) eigenvalues, it is impossible that all physical
eigenstates are $CP$ eigenstates; it is thus a sufficient condition
for indirect $CP$ violation.}

Assuming that the full
renormalized mass matrix for the neutral kaon system has  degenerate
($p^2$ dependent) eigenvalues is not unreasonable:
the basis fundamental degeneracy of the neutral kaon system keeps, at this
level, unbroken.
That the self-consistent equation defining the physical masses have
several distinct solutions is also realistic since $m^{(2)}(p^2)$ can be a
complicated function of $p^2$.
The mass splitting and indirect $CP$ violation arise
through the self-consistent procedure which mixes kinetic and
mass terms.

{\em Remark}: it can also happen that, in general $m_1^{(2)}(p^2) \not=
m_2^{(2)}(p^2)$, except at the two physical masses $\mu_1^2$ and $\mu_2^2$,
at which, consequently, $M_f(p^2=\mu_1^2)$ and $M_f(p^2=\mu_2^2)$
have  degenerate eigenvalues;
the same argumentation as above holds, though this situation looks
much less natural.

%%%%%%%%%%%%%%%%%%%%%%%%%%%%%%%%%%%%%%%%%%%%%%%%%%%%%%%%%%%%%%%%%%%%%%%%%%%%%%
\section{AN OVER-COMPLETE BASIS}
\label{section:ocbasis}
%%%%%%%%%%%%%%%%%%%%%%%%%%%%%%%%%%%%%%%%%%%%%%%%%%%%%%%%%%%%%%%%%%%%%%%%%%%%%%

A consequence of the fact that the maximal number of independent
states is two (${\cal V}_{in}\equiv {\cal V}_{out}$ is a 2-dimensional
complex vector space) is that the mass eigenstates
$|\ K_L^{in}>, |\ K_S^{in}>, |\ K_L^{out}>, |\ K_S^{out}>$ and their
hermitian conjugates are not independent
\footnote{(\ref{eq:op1}) shows that the neutral kaon
system is spanned by only two real degrees of freedom, associated respectively
with the combinations $(K^0 + e^{i\gamma}\ol{K^0})$ and $-i(K^0 -
e^{i\gamma}\ol{K^0})$.}
.

The link (\ref{eq:dual}) between states and their duals, together
with (\ref{eq:invEVM}) yields the relations
%
%\vbox{
\begin{eqnarray}
<K^0\ | = |\ K^0>^\dagger &\Rightarrow& \cr
        \frac{(1+\beta)<K_L^{out}\ |+(1+\alpha)<K_S^{out}\
|}{\sqrt{2(1-\alpha\beta)}}
&=&\frac{(1-\alpha^\ast)<K_L^{in}\ | + (1-\beta^\ast)<K_S^{in}\
|}{(\sqrt{2(1-\alpha\beta)})^\ast}
\label{eq:rel1}
\end{eqnarray}
\begin{eqnarray}
<\ol{K^0}\ | = |\ \ol{K^0}>^\dagger &\Rightarrow&\cr
 \frac{(1-\beta)<K_L^{out}\ | -(1-\alpha)<K_S^{out}\ |}
        {\sqrt{2(1-\alpha\beta)}}
 &=&\frac{(1+\alpha^\ast)<K_L^{in}\ | - (1+\beta^\ast)<K_S^{in}\ |}
        {(\sqrt{2(1-\alpha\beta)})^\ast},
\end{eqnarray}
\label{eq:rel2}
%}
%
which  exemplifies the non independence of $<K_L^{in}\ |, <K_S^{in}\ |,
<K_L^{out}\ |, <K_S^{out}\ |$.

Other relations can be  deduced from (\ref{eq:op1}) and (\ref{eq:invEVM})
and using the correspondence (\ref{eq:dual})

\vbox{
\begin{eqnarray}
<\ol{K^0}\ |=e^{-i\gamma}<K^0\ |^\dagger&& \Rightarrow\cr
\frac{(1-\beta)<K_L^{out}\ |-(1-\alpha)<K_S^{out}\ |}{\sqrt{2(1-\alpha\beta)}}
&=&
e^{-i\gamma} \frac{
(1+\beta^\ast)<K_L^{out}\ |^\dagger+(1+\alpha^\ast)<K_S^{out}\ |^\dagger}
{(\sqrt{2(1-\alpha\beta)})^\ast};\cr
&&
\label{eq:rel3}
\end{eqnarray}
}

\vbox{
\begin{eqnarray}
<{K^0}\ |=e^{-i\gamma}<\ol{K^0}\ |^\dagger&& \Rightarrow\cr
\frac{(1+\beta)<K_L^{out}\ |+(1+\alpha)<K_S^{out}\ |}{\sqrt{2(1-\alpha\beta)}}
&=&
e^{-i\gamma} \frac{
(1-\beta^\ast)<K_L^{out}\ |^\dagger-(1-\alpha^\ast)<K_S^{out}\ |^\dagger}
{(\sqrt{2(1-\alpha\beta)})^\ast}.\cr
&&
\label{eq:rel4}
\end{eqnarray}
}
Similar equations can be deduced which only involve $|\ in>$ states and
their hermitian conjugates.

Relations complementary to (\ref{eq:rel3}) and (\ref{eq:rel4}) can be obtained
from (\ref{eq:op1}) and (\ref{eq:invEVM}) without using (\ref{eq:dual}):

\vbox{
\begin{eqnarray}
<\ol{K^0}\ |=e^{-i\gamma}|\ K^0>&& \Rightarrow\cr
(1-\beta)<K_L^{out}\ |-(1-\alpha)<K_S^{out}\ | &=&
      e^{-i\gamma}\left(
    (1-\alpha)|\ K_L^{in}>+(1-\beta)|\ K_S^{in}>
              \right);\cr
&&
\label{eq:rel5}
\end{eqnarray}
}

\vbox{
\begin{eqnarray}
<K^0\ |=e^{-i\gamma}|\ \ol{K^0}>&& \Rightarrow\cr
(1+\beta)<K_L^{out}\ |+(1+\alpha)<K_S^{out}\ | &=&
      e^{-i\gamma}\left(
    (1+\alpha)|\ K_L^{in}>-(1+\beta)|\ K_S^{in}>
              \right).\cr
&&
\label{eq:rel6}
\end{eqnarray}
}
One reminds that both $|\ in>$ and $<out\ |$ states can be expanded on the
same $(K^0,\ol{K^0})$ basis, and that one can go, for example, from
$|\ K_L^{in}>$ to $<K_L^{in}\ |$ by hermitian conjugation.

(\ref{eq:rel3}) and (\ref{eq:rel4}) are more transparent than
(\ref{eq:rel5}) and (\ref{eq:rel6}) in terms of ``states'' since they only
involve $<out\ |$ states and their hermitian conjugates.

%%%%%%%%%%%%%%%%%%%%%%%%%%%%%%%%%%%%%%%%%%%%%%%%%%%%%%%%%%%%%%%%%%%%%%%%%%%%%%%
%%%%%%%%%%%%%%%%%%%%%%%%%%%%%%%%%%%%%%%%%%%%%%%%%%%%%%%%%%%%%%%%%%%%%%%%%%%%%%%
\newpage\null
\begin{em}

\end{em}

%%%%%%%%%%%%%%%%%%%%%%%%%%%%%%%%%%%%%%%%%%%%%%%%%%%%%%%%%%%%%%%%%%%%%%%%%%%%%%
\end{document}